\documentclass[11pt]{article}
\pdfoutput=1

\usepackage[a4paper,text={16.8cm,22.4cm}]{geometry}
\usepackage{amsmath,amsfonts,slashed,amssymb,tikz,bm,psfrag,graphicx,color,dsfont,euscript}
\usepackage{multicol}
\usepackage{float}
\usepackage{empheq}
\usepackage{slashed}
\usepackage{hyperref}
\usepackage[normalem]{ulem}
\RequirePackage[sort&compress,square,comma,numbers]{natbib}
\allowdisplaybreaks
\renewcommand{\arraystretch}{1.2}

\hypersetup{
  colorlinks   = true,    
  urlcolor     = blue,    
  linkcolor    = blue,    
  citecolor    = red      
}

\numberwithin{equation}{section}

\DeclareMathOperator{\Tr}{Tr}

\def\II{\hbox{{1}\kern-.25em\hbox{l}}}

\begin{document}

\begin{titlepage}

\begin{flushright}
\normalsize
 TUM-HEP-1380/21
\end{flushright}

\vspace{0.1cm}
\begin{center}
{\bf\Large $B\to P,V$ form factors with the $B$-meson  light-cone sum rules}
\end{center}
\vspace{0.5cm}

\begin{center}
{\bf  Yue-Long Shen$^a$\footnote{e-mail: shenylmeteor@ouc.edu.cn}, Yan-Bing Wei$^b$\footnote{e-mail: yanbing.wei@tum.de}}\\ \vspace{0.5cm}
{\sl $^a$ College of Physics and Photoelectric Engineering, \\Ocean University of China, Qingdao 266100,  China }\\
{\sl $^b$Physik Department T31,\\
James-Franck-Stra{\ss}e 1, Technische Universit\"at M\"unchen,\\
D–85748 Garching, Germany}\end{center}
\vspace{0.2cm}

\begin{abstract}
In this review, we discuss the calculation of the $B\to P,V$ form factors within the framework of the light-cone sum rules with the light-cone distribution amplitudes of the $B$ meson. A detailed introduction to the definition, scale evolution, and phenomenological models of the $B$-meson distribution amplitudes is presented. We show two equivalent approaches of calculating the next-to-leading order QCD corrections to the sum rules for the form factors, i.e., the method of regions and the step-by-step matching in the soft-collinear effective theory. The power suppressed corrections to the $B\to P,V$ form factors especially the contributions from the higher-twist $B$-meson distribution amplitudes are displayed. We also present numerical results of the form factors including both the QCD and the power corrections, and phenomenological applications of the predicted form factors such as the determination of the CKM matrix element $|V_{ub}|$.

\end{abstract}

\vfil

\end{titlepage}

\tableofcontents
\newpage
\section{Introduction}
The decays of $B_q$ mesons ($q = d, s$) have been playing a crucial role in the determination of the Cabibbo-Kobayashi-Maskawa (CKM) matrix elements as well as the understanding of the QCD dynamics in the heavy-light meson system. The heavy-to-light transition form factors are essential ingredients  in the semileptonic decays {$B \to M\ell \nu$ ($M$ stands for vector (V) or pseudoscalar (P) meson) }, in the flavor-changing-neutral-current (FCNC) processes $B \to M\ell^+\ell^-(\nu\bar \nu)$ and $B \to V\gamma$, and in the nonleptonic $B$-meson  decays. In the small recoil region, the heavy-to-light form factors can be determined from the experiments or calculated by the nonperturbative approach, among which the Lattice QCD simulation which based on the first principle of QCD is regarded to give the most reliable predictions. At small hadronic recoil the Lattice QCD calculations of $B \to \pi, K$, $B_s \to K$ form factors have been performed
\cite{Lattice:2015tia,Bailey:2015nbd,Bailey:2015dka} using the gauge-field ensembles with (2+1)-flavour lattice configurations. 
In addition, the Flavour Lattice Average Group (FLAG) has given the results of these form factors with an extrapolation to the whole kinematic region from the small hadronic recoil region of the light meson \cite{FlavourLatticeAveragingGroup:2019iem}. The unquenched lattice QCD calculations of $B \to K^{\ast}$
form factors have been performed \cite{Horgan:2013hoa,Horgan:2015vla} by employing the gauge-field ensembles with an improved staggered quark action from the MILC Collaboration \cite{Bazavov:2009bb}.

Due to the limited computing capability, the Lattice simulation cannot be applied to the large recoil region directly. In the framework the QCD factorization, the heavy-to-light form factors at large recoil contain both the soft contribution satisfying the large-recoil symmetry relations and the hard spectator scattering effect violating the symmetry relations at leading power in $\Lambda/m_{Q}$\cite{Beneke:2000wa}.
The soft-collinear effective theory (SCET) provides a more transparent insight on the factorization property of heavy-to-light form factors
 by integrating out the hard and hard-collinear fluctuations step by step.
Implementing the first-step matching procedure for the QCD current $\bar \psi \, \Gamma_i \, Q$ will give rise to the so-called
${\rm A0}$-type and ${\rm B1}$-type ${\rm SCET}_{\rm I}$  operators \cite{Bauer:2002aj,Beneke:2002ph,Beneke:2003pa}, both of which can contribute to heavy-to-light form factors at leading power  in $\Lambda/m_{Q}$. The matrix elements of the ${\rm A0}$-type operator are non-factorizable due to the emergence of endpoint divergences in the
convolution integrals of the jet functions from the matching between ${\rm SCET}_{\rm I}$ and ${\rm SCET}_{\rm II}$ and the light-cone distribution amplitudes (LCDA).
By contrast, the matrix elements of the ${\rm B1}$-type ${\rm SCET}_{\rm I}$ operators can be further factorized into
 convolutions of the jet functions and the LCDAs \cite{Beneke:2003pa}. Since the latter one is suppressed by the strong coupling constant, the heavy-to-light form factors are dominated by the soft form factor in the QCD factorization.
An alternative approach to compute the heavy-to-light form factors is based upon the transverse-momentum-dependent
(TMD)  factorization for hard processes,  where the on-shell Sudakov form factor \cite{Collins:1989bt} arises from the resummation of large logarithmic terms that can effectively suppress the region with small transverse momentum \cite{Botts:1989kf}. In the TMD factorization approach, which is also called the PQCD approach, the endpoint singularity will disappear, and then the form factors are perturbative calculable.  The $B \to \pi$ form factors within the PQCD approach have been pushed to ${\cal O}(\alpha_s)$ for twist-2 \cite{Li:2012nk,Li:2012md}
and twist-3 \cite{Cheng:2014fwa} contributions of pion LCDAs. However,  the infrared subtractions beyond the leading order in $\alpha_s$ \cite{Li:2014xda} are much more complex than that in the QCD factorization and a complete understanding of the TMD factorization for exclusive processes with large momentum transfer has not been achieved to date on the conceptual side.

The Lattice QCD simulation for the heavy-to-light form factors is valid in the small recoil region, its predictions need to be extrapolated to the whole kinematic region from the small hadronic recoil of the light meson through special models. The extrapolation-model dependence will produce unavoidable uncertainties in the determination of observations, especially at large recoil. Therefore, to obtain the $q^2$ (invariant mass of lepton pair in the semi-leptonic decays) dependence of the form factors with high accuracy, it is important to compute the form factors in the large hadronic recoil region directly. The light-cone sum rules (LCSR) approach, which is a combination of the SVZ sum rules with the QCD theory of hard exclusive processes, provides an appropriate method to evaluate the form factors at large recoil.  According to the correlation functions employed in the calculation,  two different frameworks of LCSR, i.e., LCSR with the light-meson LCDAs \cite{Khodjamirian:2017fxg,Ball:2001fp,Khodjamirian:2011ub,Ball:1997rj,Ball:1998kk,Ball:2004rg,Straub:2015ica} and   LCSR with the $B$-meson LCDAs (we will also call this the $B$-meson LCSR) \cite{Khodjamirian:2005ea,Khodjamirian:2006st,Khodjamirian:2010vf,Khodjamirian:2012rm}, have been established. The advantage of LCSR with light-meson LCDAs is that it can be applied to a larger region of the square of the momentum transfer and also the LCDAs of the light meson are better determined than that of the $B$ meson. Meanwhile, the $B$-meson LCSR has its unique advantage that the input LCDAs are universal for all the $B\to M$ form factors, and the theoretical uncertainty can be sizeably reduced in the calculation of ratios of the form factors. 
The LCSR for the ${\rm SCET}_{\rm I}$ matrix elements entering the QCD factorization formulae of
heavy-to-light $B$-meson decay form factors, which is also called the SCET sum rules, has been achieved \cite{DeFazio:2005dx,DeFazio:2007hw}
employing the vacuum-to-$B$-meson correlation functions. To improve the accuracy of the $B\to P~(P=\pi,K,D)$ form factors, the next-to-leading order (NLO) corrections to the correlation functions in the strong coupling $\alpha_s$ have been performed in \cite{Wang:2015vgv,Shen:2016hyv,Wang:2017jow,Lu:2018cfc} ,
where the factorization formulae for the correlation functions were established with the
diagrammatic approach and the strategy of regions \cite{Beneke:1997zp,Smirnov:2002pj}. With a similar method, the baryonic transition $\Lambda_b \to \Lambda$ form factors have also been investigated \cite{Wang:2015ndk}.
The $B \to V$ form factors have also been evaluated with the  SCET sum rules and the NLO correction to the hard-collinear function is performed \cite{Gao:2019lta}.

The fundamental nonperturbative inputs in the $B$-meson LCSR are the LCDAs of the $B$ meson. 
Because they are nonperturbative the hard effect must be integrated out in the definition of the LCDAs. Therefore, one has to employ the effective bottom quark field in the heavy quark effective theory (HQET) to construct the matrix elements of the nonlocal operators \cite{Grozin:1996pq}. As a soft objective the $B$-meson LCDAs do not have a definite twist in principle, while when entering a process with large momentum transfer and thus only one of the light-cone components of the momentum of the soft light quark would be picked up, the twist of the $B$-meson LCDAs makes sense. The scale dependence of the LCDAs of the $B$ meson can be obtained by the renormalization group (RG) equation approach, and the  RG equation for the leading-twist LCDA has been derived at two-loop level \cite{Lange:2003ff,Braun:2019wyx}, and the one-loop-level evolution equations for higher-twist LCDAs are also known \cite{Braun:2015pha}. The evolution behavior of the LCDAs can provide constraints on the model of the LCDAs, in addition, it also plays an important role in proving the factorization-scale independence of the factorization formula.

The plan of this review is as follows. In Section \ref{sec:lcda} we will discuss the LCDAs of the $B$ meson, including the definition, the evolution behavior, and the phenomenological models of the leading-twist and higher-twist LCDAs. In Sections \ref{sec:lcsr} and \ref{sec:nlo} we will introduce the LCSR with $B$-meson LCDAs, emphasizing the two equivalent methods to evaluate the QCD corrections to the correlation functions, i.e., the method of regions and the matching between the $\rm SCET_I$ and $\rm SCET_{II}$. In Section \ref{sec:nlp} we will discuss the power suppressed contributions to the heavy-to-light form factors with LCSR, concentrating on the contributions from high-twist $B$-meson LCDAs. In Section \ref{sec:results} we will present some numerical results of the form factors and the phenomenological applications. We summarize in the last section.

\section{The light-cone distribution amplitudes of the $B$-meson}
\label{sec:lcda}

\subsection{Definition of the leading-twist $B$-meson LCDAs}

The $B$-meson LCDAs are among the most important ingredients of the QCD factorization formula for exclusive $B$ decays.
Before talking about the heavy-to-light form factors, we would like to introduce the LCDAs of the $B$ meson.
The two-particle LCDAs of the $B$ meson in the HQET  can be obtained from the coordinate-space matrix elements \cite{Grozin:1996pq}
\begin{eqnarray}\langle0|\bar q^\beta(z)[z,0]h_v^\alpha(0)|\bar B(v)\rangle=-{i\tilde f_B(\mu) m_B\over 4}
\left[{1+\not\!v\over 2}\left\{{ 2}\Phi_+{(t,z^2)}+{\Phi_-{(t,z^2)}-\Phi_+{(t,z^2)}\over t}\not\!z\right\}\right]^{\alpha\beta}\, ,
\end{eqnarray}
where $t=v\cdot z$.
The LCDAs $\Phi_\pm{(t,z^2)}$  can be expanded around $z^2= 0$. In the limit $z^2\to 0$, we assume $t\to \tau={ n\cdot z/2}$, then the $B$-meson LCDAs in the momentum space are defined through the Fourier transformation
\begin{eqnarray}
\phi_{\pm}(\omega)=\int {d\tau\over 2\pi} \, e^{i\omega \tau} \, \Phi_{\pm}({\tau})\,,
\qquad
\Phi_{\pm}({\tau}) \equiv  \Phi_{\pm}({\tau},0).
\end{eqnarray}
For convenience, we introduce two light-like vectors $n$ and $\bar n$ satisfying $n\cdot \bar n=2, n^2=\bar n^2=0$.
Any four-vector $k^\mu$ can be expressed as
$k^\mu = n \cdot k\, \bar n^\mu/2 + \bar n \cdot k\, n^\mu/2 + k^\mu_\perp$.

When the LCDAs of the $B$ meson are applied in the calculation of $B$-meson decay processes, the momentum space projector is usually required. To obtain the projector, we adopt the reference frame satisfying $v=(n+\bar n)/2$ and assume $\bar n\cdot z\ll z_\perp \ll n\cdot z$, thus $\not\! z\simeq \tau \not\! \bar n+\not\! z_\perp$, then
\begin{eqnarray}
\Phi_+{(t,z^2)}+{\Phi_-{(t,z^2)}-\Phi_+{(t,z^2)}\over t}\not\!z
\simeq
\Phi_+{(\tau)}\not\! n+\Phi_-{(\tau)}\not\! \bar n+{\Phi_-{(\tau)}-\Phi_+{(\tau)}\over \tau}\not\! z_\perp.
\end{eqnarray}
After the Fourier transform, the projector of $B$-meson LCDAs is obtained as \cite{Beneke:2000wa}
\begin{eqnarray}
{\cal M}_B(\omega)=-{i\tilde f_B(\mu) m_B\over 4}
\left[{1+\not\!v\over 2}\left\{\phi_+(\omega)\not\! n
+\phi_-(\omega)\not\! \bar n
-\int_0^\omega [{\phi_-{(\omega)}-\phi_+{(\omega)}}]\gamma_\perp^\mu {\partial \over \partial k^\mu_\perp}\right\}\right]\, ,
\end{eqnarray}
with $k$ is the momentum of the anti-light quark in the $B$-meson.

The LCDAs $\Phi_\pm$ defined above do not contribute at the same power in the QCD factorizations of the heavy hadron decay processes, since they do not have the same collinear twist (In the following we call ``twist" for short). The twist $t$ and conformal spin $j$ of the light quark and gluon fields are given by the usual expressions \cite{Braun:2003rp}
\begin{align}
   t = d - s\,,  \qquad j = \frac12(d+s)\,,
\end{align}
where $d$ is the canonical dimension and $s$ is the spin {projection} on the light-cone. The twist and conform spin are closely related to the collinear subgroup of the conformal group.  The collinear subgroup of the conformal group is locally equivalent to the $SL(2,R)$ group, and it contains four generators, namely $P_+, M_{-+}, D$ and $K_-$, where $P_+$ and $M_{-+}$ are the projections of the generator of the Poincare group $p_\mu, M_{\nu\mu}$ on the light cone, $D$ is the generator of dilatation, and $K_-$ is the generator of the special conformal transformation along the light-cone \cite{Braun:2003rp}. Specifically, if we write the light-like vector $x=\alpha n$, then the dilatation indicates the transform $\alpha \to \alpha'=\lambda \alpha$, and for the special conformal transformation,
\begin{align}
  \alpha \to \alpha'={a\alpha+b\over c\alpha +d},\,\,\,ad-bc=1.
\end{align}
The four generators of the collinear subgroup of the conformal group can be rearranged to form the algebra of $SL(2,R)$, i.e.,
\begin{eqnarray}
  { J}_+ = { J}_1 + i { J}_2  = - i { P}_+\,, &~~~~~~&
  { J}_- = { J}_1 - i { J}_2  = ({i}/{2}) { K}_-\,,
\nonumber\\
 { J}_0 = ({i}/{2}) ({ D}+{ M}_{-+})\,,  &~~~~~~&
 { E} =   ({i}/{2})({ D}-{ M}_{-+})\,.
\end{eqnarray}
with
\begin{eqnarray}
  [{ J}_0,{ J}_\mp] =\mp  {J}_\mp\,, &~~&  [{ J}_-,{ J}_+] = -2 { J}_0 \,,
\end{eqnarray}
For the quantized field $\Phi(\alpha)$, we have
\begin{eqnarray}
 {}[{ J}_+,\Phi(\alpha)] &=& - \partial_\alpha\Phi(\alpha) ,\,\,\,
 {}[{ J}_-,\Phi(\alpha)] =  \left(\alpha^2\partial_\alpha+2 j \alpha \right)\Phi(\alpha),
          \nonumber\\
 {}[{ J}_0,\Phi(\alpha)] &=&  \left(\alpha\partial_\alpha+ j\right)\Phi(\alpha),\,\,\,[{\bf E},\Phi(\alpha)] = \frac12(\ell-s)\Phi(\alpha)\,.
\end{eqnarray}
For a spinor field $\Psi$, the projectors $\Psi_+={\not n \not \bar n\over 4}\Psi$ and $\Psi_-={\not\bar n \not n\over 4}\Psi$ have definite twist, namely $t[\Psi_\pm]=\pm1$. In an appropriate reference frame, $\Psi_\pm$ only have two nonzero component, thus it is more convenient to define these fields with two-component spinors. Any light-like vector can be represented by a product of two spinors. One can write
\begin{align}
n_{\alpha\dot\alpha}=n_\mu\sigma^\mu_{\alpha\dot\alpha}=\lambda_\alpha\bar\lambda_{\dot\alpha}\,,
&& \bar n_{\alpha\dot\alpha}=\bar n_\mu\sigma^\mu_{\alpha\dot\alpha}=\mu_\alpha\bar\mu_{\dot\alpha}
\end{align}
where for the auxiliary $\lambda$ and $\mu$ spinors, $\bar\lambda = \lambda^\dagger$, $\bar\mu = \mu^\dagger$ which satisfy
$
(\lambda\,\mu)=\lambda^\alpha \mu_\alpha=2$,
$(\bar\mu\, \bar\lambda)=\bar\mu_{\dot\alpha} \bar\lambda^{\dot\alpha}=2
$. The ``+'' and ``--'' fields are defined as,
\begin{align}
\chi_+=\lambda^\alpha \psi_\alpha, && \bar\psi_+=\bar\lambda^{\dot \alpha} \psi_{\dot\alpha},
&& f_{++}=\lambda^\alpha\lambda^\beta f_{\alpha\beta}, && f_{+-}=\lambda^\alpha\mu^\beta f_{\alpha\beta}\,,
&&
\bar f_{++}=\bar\lambda^{\dot \alpha}\bar\lambda^{\dot\beta} \bar f_{\dot\alpha\dot\beta}
\end{align}
etc.
 The Dirac
(antiquark) spinors
$$
 q=\begin{pmatrix}
\psi_\alpha\\\bar \chi^{\dot\beta}\end{pmatrix}\,, \qquad\qquad \bar q=(\chi^\beta,\bar\psi_{\dot\alpha})
$$
are written in term of the following two component fields
\begin{align}
(\lambda\mu)\,\chi^\alpha=\mu^\alpha \chi_+  - \lambda^\alpha \chi_-\,,
&&
(\bar \mu\bar\lambda)\,\bar \psi_{\dot\alpha}=\bar \mu_{\dot\alpha} \bar \psi_+  - \bar\lambda_{\dot\alpha}\bar \psi_-\,.
\end{align}
The large component of the heavy quark field in the  HQET satisfies $\slashed{v}h_v = h_v $ , then
\begin{align}
    h_{+} = - \bar h_-\,, \qquad h_- =  \bar h_+\,.
\label{EOMheavy}
\end{align}
The gluon strength tensor $F_{\mu\nu}$ can be decomposed as
\begin{align}
F_{\alpha\beta,\dot\alpha\dot\beta} =\sigma^\mu_{\alpha\dot\alpha} \sigma^\nu_{\beta\dot\beta} F_{\mu\nu}=
2\left(\epsilon_{\dot\alpha\dot\beta} f_{\alpha\beta}-
\epsilon_{\alpha\beta} \bar f_{\dot\alpha\dot\beta}
\right),
\notag\\
i {\widetilde F}_{\alpha\beta,\dot\alpha\dot\beta}=
\sigma^\mu_{\alpha\dot\alpha} \sigma^\nu_{\beta\dot\beta}i\widetilde F_{\mu\nu}=
2\left(\epsilon_{\dot\alpha\dot\beta}f_{\alpha\beta}+
\epsilon_{\alpha\beta}\bar f_{\dot\alpha\dot\beta}\right).
\end{align}
Here $f_{\alpha\beta}$ and $\bar f_{\dot\alpha\dot\beta}$ are chiral and antichiral
symmetric tensors, $f^*=\bar f$, which belong to $(1,0)$ and $(0,1)$ representations
of the Lorentz group, respectively. It is easy to see that the twist of the field $\psi_+$ and $\chi_+$ is 1, the twist of $\psi_-$ and $\chi_-$ is 2, and we assign the twist of $h_\pm$ to be 1. Then the relevant operators with definite twist in spinor notation one obtains 
\begin{align}
\tilde f_B(\mu) m_B \Phi_+(z;\mu) & = i\,\langle 0| \bar\psi_+(z) h_+(0) - \chi_+(z) \bar h_+(0)|\bar B(v)\rangle\,,
\notag\\[2mm]
\tilde f_B(\mu) m_B \Phi_-(z;\mu) & = i\,\langle 0| \bar\psi_-(z) h_-(0) - \chi_-(z) \bar h_-(0)|\bar B(v)\rangle\,,
\end{align}

\subsection{Evolution of leading twist $B$-meson LCDA}

At leading power, only $\phi_{+}(\omega)$ is relevant in the factorization formula of various $B$-meson decay processes, and the evolution equation of $\phi_{+}(\omega)$ is the well-known Lange-Neubert equation \cite{Lange:2003ff}, which reads:
\begin{eqnarray}
\frac { d } { d \ln \mu } \phi_ { + } ( \omega ,\mu ) &=&
-  \int _ { 0 } ^ {
\infty } d \omega ^ { \prime }\Gamma_+(\omega,\omega',\mu)\phi_{+} \left( \omega ^ {
\prime } , \mu \right),\nonumber \\
\Gamma_+(\omega,\omega',\mu) &=& \left( \Gamma_{\rm cusp} \ln
\frac { \mu} { \omega } + \gamma \right) \delta \left( \omega -
\omega ^ { \prime } \right) + \omega \, \Gamma_{\rm cusp}\,\Gamma(\omega,\omega')\, ,
\label{LN:evolution:eq}
\end{eqnarray}
where $\mu$ is the renormalization scale. At the one-loop level, the
anomalous dimensions are
\begin{align}
\Gamma_{\rm cusp}= &~\frac{\alpha_{s}C_F}{\pi},
&\gamma^{(0)}=&-\frac{\alpha_{s}C_F}{2\pi},
&\Gamma(\omega,\omega') =&- \left[ \frac { \theta \left( \omega ^
{ \prime } - \omega \right) } { \omega^{ \prime } \left( \omega ^
{ \prime } - \omega \right) }+\frac { \theta \left( \omega -
\omega ^ { \prime } \right) } { \omega \left( \omega - \omega ^ {
\prime } \right) } \right] _ { + }\,,
\end{align}
with the ``plus" function is defined as
\begin{eqnarray}
\int^\infty_0 dy \,\Big [ f(x,y) \Big ]_{+} g(y)=
\int^\infty_0 dy \, f(x,y) \,\Big[ g(y)-g(x) \Big]\,.
\end{eqnarray}
Since the evolution equation of the leading twist $B$-meson LCDA is the integro-differential equation, it is difficult to obtain the solution directly. A commonly used method is to simplify the evolution equation by an integral transformation.  There exists several kinds of integral transform  which are helpful to work out the solution of the evolution equation.
It was found that the evolution kernel is diagonalized  when it is transformed into the so-called ``dual" space \cite{Bell:2013tfa}. The leading twist LCDA in the dual space can be obtained by
\begin{eqnarray}\label{dual}
\rho_{+}\left(\omega^{\prime}, \mu\right)=\int_{0}^{\infty} \frac{d \omega}{\omega}
\sqrt{\frac{\omega}{\omega^{\prime}}} J_{1}\left(2 \sqrt{\frac{\omega}{\omega^{\prime}}}\right)
\phi_{+}(\omega, \mu),
\end{eqnarray}
which satisfies an ordinary differential equation:
\begin{eqnarray}
\mu \, {d \over d \mu} \,\rho_{+}(\omega',\mu) = -\Big[\Gamma_{\rm
cusp}\ln{\mu\over e^{-2 \gamma_{E}} \omega'}+\gamma \Big]\rho_{+} (\omega',\mu)\, .
\end{eqnarray}

The evolution equation can also be calculated in the position space, where it takes the form \cite{Kawamura:2010tj}
\begin{eqnarray}
\frac { d } { d \ln \mu } \Phi_{ + } ( t ,\mu ) &=&
- \big[\Gamma_{\rm cusp}(\alpha_s)\ln{it\tilde\mu}+\gamma_+(\alpha_s)-\gamma_F(\alpha_s)\big]\,\Phi_{ + } ( t ,\mu )+\int _ { 0 } ^ {
1} d z \,K(z,\alpha_s)\,\Phi_{ + } ( zt ,\mu ) \, .
\label{positionspace}
\end{eqnarray}
This equation can be related to the LN equation by a  Fourier transform.
Performing the Mellin transformation to the evolution equation in the position space \cite{Kawamura:2010tj,Braun:2019zhp}:
\begin{align}
\tilde\varphi_{+}(j,\mu) =&~ \frac{1}{2\pi i} \int^{-i\infty}_{-i0} \frac{dt}{t} \,
(it\tilde \mu )^{-j} \,
\Phi_{+}(t,\mu) \,,
\end{align}
the evolution equation can also be diagonalized and easily  solved. The different kinds of integral transform mentioned above are equivalent, and the LCDAs $\phi_{+}(\omega)$, $~\Phi_{+}(t)$,$ ~\tilde \varphi_{+}(j)$,$~ \rho_{+}(\omega')$  are different expressions of an identical objective.
Because the momentum space and the position space are related through a standard Fourier transformation, we are able to derive
\begin{eqnarray}\label{MellMom}
\tilde\varphi_+(j) &=& \frac{\Gamma(-j)}{2\pi i} \, \int^{\infty}_{0} d\omega\,
\Big(\frac{\omega}{\tilde\mu }\Big)^j \, \phi_+(\omega) \,,
\nonumber \\
\tilde\varphi_{+}(j,\mu)&=&{\tilde \mu\over 2\pi i} \, {\Gamma(2+j) }\int_0^\infty {d\omega'\over \omega'}\rho_{+}(\omega',\mu)\left({\tilde\mu\over \omega'}\right)^{-1-j}\, .
\end{eqnarray}
At the one-loop level, the most convenient method  is to work in the dual space since the Bessel function is the eigenfunction of the Lange-Neubert kernel, which is confirmed in \cite{Braun:2014owa,Braun:2018fiz}. The Lange-Neubert kernel can be expressed as a logarithm of the generator of special conformal transformations along the light cone. When the eigenfunction of the generator is transformed to the momentum space, it is simply the Bessel function in Eq.(\ref{dual}).

The two-loop-level anomalous dimension of the $B$-meson LCDA was first calculated in the coordinate space in \cite{Braun:2019wyx}, and it is more simply expressed in the dual space:
\begin{eqnarray}
\left[\mu{\partial\over \partial \mu} +\beta(a){\partial\over \partial{a}}+\Gamma_{\rm cusp}(a)\ln(\tilde \mu e^{\gamma_E}s)+\gamma(a)\right]\eta_+(s,\mu)=4\,C_F\,a^2
\int_0^1{du\over u} \,\bar u \,h(u)\,\eta_+(\bar us,\mu)\, ,
\end{eqnarray}
where $s\eta_+(s)=\rho_{+}(1/s)$ and $a=\alpha_s/(4\pi)$. This equation has also been transformed into the momentum space in \cite{Liu:2020ydl}, resulting in the two-loop-level Lange-Neubert equation. The advantage of solving the evolution equation at the two-loop level in the dual space does not hold since the two-loop evolution kernel is not diagonal in this space. On the contrary, the elegant form of the evolution equation in the Mellin space  is maintained
\begin{align}\label{mellinequation}
\Big[\frac{d}{d\ln\mu}
+\hat V(j,\alpha_s)\Big]\,\tilde\varphi_{+}(j,\mu)
=0\,,
\end{align}
with
\begin{align}
\hat V(j,\alpha_s)
=j+\gamma_+-\gamma_F+
\Gamma_{\rm cusp}\,\Big[\psi(j+2)-\psi(2)+\vartheta(j)\Big]\,,
\end{align}
where \begin{align}
  \vartheta(j) &= a  \vartheta^{(1)}(j) = a \biggl\{(\beta_0-3C_F)\Big(\psi^\prime(j+2)-\psi^\prime(2)\Big)
+2C_F\biggl(\frac1{(j+1)^3}
 \notag\\
&\quad
+\psi^\prime(j+2)(\psi(j+2)-\psi(1))+\psi^\prime(j+1)(\psi(j+1)-\psi(1))-\frac{\pi^2}6\biggr)
  \biggr\},
  \notag\\
  \gamma_+(a) &=
   -a C_F +  a^2 C_F
\biggl\{
4 C_F \left[\frac{21}{8} + \frac{\pi^2}{3} - 6\zeta_3\right]
+ C_A \left[\frac{83}{9} -\frac{2\pi^2}{3} - 6\zeta_3\right]
+ \beta_0\left[\frac{35}{18} -\frac{\pi^2}{6}\right]
\biggr\}\,,
\nonumber \\
\gamma_F(a) &= -3 a C_F + a^2 C_F\bigg\{C_F\Big[\frac{5}{2}-\frac{8\pi^2}{3}\Big]
+ C_A \Big[1+ \frac{2\pi^2}{3}\Big] -\frac{5}{2} \beta_0\bigg\}.
\end{align}
The solution in the Mellin space is then obtained directly \cite{Braun:2019zhp}
\begin{align}\label{mellinsolution}
\tilde\varphi_{+}(j(\mu),\alpha_s(\mu),\mu)=\tilde\varphi_{+}(j(\mu_0),\alpha_s(\mu_0),\mu_0)\exp
\left\{-\int_{\mu_0}^\mu{ds\over s}\hat V[j(s),\alpha_s(s)]\right\}\, .
\end{align}

In a recent paper \cite{Galda:2020epp}, an alternative approach to solving the evolution equation at the two-loop level was proposed. The essential idea of this approach is to perform a Laplace transformation on the $B$-meson LCDA,
\begin{align}
\tilde\phi_{+}(\eta,\mu) = \int^\infty_0 \frac{d\omega}{\omega}\,
\Big(\frac{\omega}{\bar \omega}\Big)^{-\eta} \, \phi_{+}(\omega,\mu) \,,
\end{align}
where $\bar\omega$ is a fixed reference scale, which can be used to eliminate the logarithmic moment $\sigma_1$ in the factorization formula of $B \to \gamma\ell\bar \nu_{\ell}$.
We note that the LCDA $ \varphi_{+}(j)$ is related to $\tilde\phi_{+}(\eta)$ through \cite{Shen:2020hfq}
\begin{align}
\tilde\varphi_{+}(j,\mu)
=&~\frac{\Gamma(-j)}{2\pi i} \,e^{\gamma_{E}} \mu\,
\Big(\frac{\bar \omega}{e^{\gamma_{E}} \mu}\Big)^{j+1}\, \,
\tilde\phi_{+}(-j-1,\mu) \,.
\end{align}
Then, one could derive the RG equation for $\tilde\phi_{+}$ and solve this equation directly \cite{Galda:2020epp,Shen:2020hfq}.

\subsection{High twist $B$-meson LCDAs}
Power corrections in the $B$-meson decay processes are of great importance, and higher twist LCDAs provide a kind of important power suppressed contributions. The LCDA $\phi_-$ defined in the previous subsection is twist-3, and it is suppressed due to different component of the quark field in the definition of LCDAs. Besides, the additional gluon or quark fields will also give rise to high-twist LCDAs. Compared with two-particle LCDAs, the three-particle quark-gluon LCDAs are more numerous. There exist eight independent Lorentz structures~\cite{Geyer:2005fb} and they can be defined as
\begin{eqnarray}
\lefteqn{\langle 0| \bar q(nz_1) gG_{\mu\nu}(nz_2)\Gamma h_v(0) |\bar B(v)\rangle =}
\nonumber\\
&=&
\frac12 \tilde f_B(\mu) m_B \Tr\biggl\{\gamma_5 \Gamma P_+
\biggl[ (v_\mu\gamma_\nu-v_\nu\gamma_\mu)  \big[{\Psi}_A-{\Psi}_V \big]-i\sigma_{\mu\nu}{\Psi}_V
- (n_\mu v_\nu-n_\nu v_\mu){X}_A
\nonumber\\&&{}\hspace*{0.1cm}
 + (n_\mu \gamma_\nu-n_\nu \gamma_\mu)\big[W+{Y}_A\big]
- i\epsilon_{\mu\nu\alpha\beta} n^\alpha v^\beta \gamma_5 \widetilde{X}_A
+ i\epsilon_{\mu\nu\alpha\beta} n^\alpha \gamma^\beta\gamma_5 \widetilde{Y}_A
\nonumber\\&&{}\hspace*{0.1cm}
- (n_\mu v_\nu-n_\nu v_\mu)\slashed{n}\,{W} + (n_\mu \gamma_\nu-n_\nu \gamma_\mu)\slashed{n}\,{Z}
\biggr]\biggr\}(z_1,z_2;\mu)\,.
\label{def:three}
\end{eqnarray}
where the totally antisymmetric tensor with $\epsilon_{0123} =1$, the covariant derivative is defined as
$D_\mu = \partial_\mu -igA_\mu$ and the dual gluon strength tensor as
$\widetilde{G}_{\mu\nu} = \frac12 \epsilon_{\mu\nu\alpha\beta} G^{\alpha\beta}$.
The momentum space distributions are defined through Fourier transforms
\begin{align}
 {\Psi}_A (z_1,z_2) &=
\int_0^\infty \!\!d\omega_1\!  \int_0^\infty \!\!d\omega_2\,\, e^{-i\omega_1 z_1-i\omega_2 z_2}\, {\psi}_A (\omega_1,\omega_2)
\label{3pt-momspace}
\end{align}
and similarly for the other functions. The LCDAs defined above do not have definite twist. In order to construct the LCDAs with definite twist, one can take advantage of the two-component spinors as follows
\begin{align}
   {2} \tilde f_B(\mu) m_B \Phi_3(z_1,z_2;\mu) &=-
\langle 0|\chi_+(z_1)\bar f_{++}(z_2)  h_+(0) + \bar\psi_+(z_1) f_{++}(z_2) \bar h_+(0) |\bar B(v)\rangle\,,\nonumber \\
    2 \tilde f_B(\mu) m_B \Phi_4(z_1,z_2;\mu) &=
\phantom{-}  \langle 0|\chi_-(z_1)  f_{++}(z_2) h_-(0) + \bar\psi_-(z_1) \bar f_{++}(z_2) \bar h_-(0) |\bar B(v)\rangle\,,
\notag\\[2mm]
   \tilde f_B(\mu) m_B \big[ \Psi_4 + \widetilde{\Psi}_4\big](z_1,z_2;\mu) &=
    - \langle 0| \chi_+(z_1)f_{+-}(z_2) h_-(0) +  \bar\psi_+(z_1)\bar f_{+-}(z_2) \bar h_-(0) |\bar B(v)\rangle\,,
\notag\\[2mm]
   \tilde f_B(\mu) m_B \big[ \Psi_4 - \widetilde{\Psi}_4\big](z_1,z_2;\mu) &=
    - \langle 0| \chi_+(z_1) \bar f_{+-}(z_2)h_-(0) + \bar\psi_+(z_1) f_{+-}(z_2) \bar h_-(0)|\bar B(v)\rangle\,
    \\
     2 \tilde f_B(\mu) m_B \Phi_5(z_1,z_2;\mu) &=
  \langle 0|\chi_+(z_1)  f_{--}(z_2) h_+(0) + \bar\psi_+(z_1)\bar  f_{--}(z_2) \bar h_+(0) |\bar B(v)\rangle\,,
\notag\\[2mm]
   \tilde f_B(\mu) m_B \big[ \Psi_5 + \widetilde{\Psi}_5\big](z_1,z_2;\mu) &=
   \langle 0| \chi_-(z_1)f_{+-}(z_2)h_+(0) + \bar\psi_-(z_1)\bar f_{+-}(z_2) \bar h_+(0) |\bar B(v)\rangle\,,
\notag\\[2mm]
   \tilde f_B(\mu) m_B \big[ \Psi_5 - \widetilde{\Psi}_5\big](z_1,z_2;\mu) &=
    \langle 0| \chi_-(z_1) \bar f_{+-}(z_2)h_+(0) + \bar\psi_-(z_1) f_{+-}(z_2) \bar h_+(0)|\bar B(v)\rangle\,,\\
    {2} \tilde f_B(\mu) m_B \Phi_6(z_1,z_2;\mu) &=~
\langle 0| \chi_-(z_1)\bar f_{--}(z_2)  h_-(0) + \bar\psi_-(z_1) f_{--}(z_2) \bar h_-(0) |\bar B(v)\rangle\,..
\label{spinor4}
\end{align}
This eight invariant function is related to the $B_q$-meson higher-twist LCDAs:
\begin{equation}
\begin{aligned}
		& \Phi_3 = \Psi_A - \Psi_V, \hspace{2.5cm} \Phi_4 = \Psi_A + \Psi_V.
		\\
		& \Psi_4 = \Psi_A + X_V, \hspace{2.5cm} \tilde \Psi_4 = \Phi_A - \tilde X_A.
		\\
		& \Psi_5 = -\Psi_A + X_A - 2Y_A, \hspace{1cm} \Phi_5 = \Psi_A + \Psi_V + 2Y_A - 2\tilde Y_A + 2W.
		\\
		& \tilde \Psi_5 = -\Psi_V - \tilde X_A + 2\tilde Y_A, \hspace{1cm}	\Phi_6 = \Psi_A - \Psi_V + 2Y_A + 2\tilde Y_A + 2W - 4Z.
\end{aligned}  \end{equation}

Except for the higher fock state,  the higher twist also arises from the nonvanishing parton transverse momenta (or virtuality). The twist-4 and twist-5 two-particle $B$-meson LCDAs can be defined as
\begin{eqnarray}
 \langle 0| \bar q(x) \Gamma [x,0] h_v(0) |\bar B(v)\rangle &=&
-\frac{i}2 \tilde f_B(\mu) m_B \Tr\Big[\gamma_5 \Gamma P_+ \Big] \int\limits_0^\infty d\omega \, e^{-i\omega (vx)}
\Big\{\phi_+(\omega) + x^2 g_+(\omega)\Big\}
\nonumber\\&&{}\hspace*{-4cm} +\frac{i}4  \tilde f_B(\mu) m_B \Tr\Big[\gamma_5 \Gamma P_+ \slashed{x} \Big] \frac{1}{vx}
\int\limits_0^\infty d\omega \, e^{-i\omega (vx)} \Big\{[\phi_+-\phi_-](\omega) + x^2 [g_+-g_-](\omega)\Big\}
\label{def:g+g-}
\end{eqnarray}
where we have assuming $|x^2| \ll 1/\Lambda_{\rm QCD}^2$, thus Eq.~\eqref{def:g+g-} can  be understood as a light-cone expansion to the tree-level accuracy. This definition contains the constraints
\begin{align}
   \int\limits_0^\infty d\omega \,  \Big[\phi_+(\omega) -\phi_-(\omega)\Big] = 0\,,
&&
   \int\limits_0^\infty d\omega \,  \Big[g_+(\omega) -g_-(\omega)\Big] = 0\,.
\end{align}

From QCD equation of motion, on can derive the following relations between diffrerent LCDAs~\cite{Kawamura:2001jm}%
\footnote{The last two relations in~\eqref{KKQT} follow from the expressions given in~\cite{Kawamura:2001jm} by simple algebra.}
\begin{subequations}
\label{KKQT}
\begin{align}
\hspace*{-0.5cm}  \Big[z\frac{d}{dz}+1\Big]\Phi_-(z) &=  \Phi_+(z)  + 2 z^2  \int_0^1\! udu\,\Phi_3(z,uz)\,,
\label{KKQT1}
\\
2 z^2  \mathrm{G}_+(z) & =
-  \Big[ z \frac{d}{dz} - \frac12  + i z \bar \Lambda \Big] \Phi_+(z)
-  \frac{1}{2}\Phi_-(z)
- z^2  \int_0^1\! \bar udu\,{\Psi}_4(z,uz)\,,
\label{KKQT2}
\\
 2 z^2 \mathrm{G}_-(z)
&= -  \Big[ z \frac{d}{dz} - \frac12  + i z \bar \Lambda \Big] \Phi_-(z) - \frac12  \Phi_+(z)
- z^2  \int_0^1\! \bar udu\,{\Psi}_5(z,uz)\,,
\label{KKQT3}
\\
 \Phi_-(z)
&= \left(z \frac{d}{dz}+1 + 2i z \bar \Lambda  \right) \Phi_+(z) +
2 z^2 \int_0^1\! du\,  \Big[ u \Phi_4(z,uz) + {\Psi}_4(z,uz)\Big],
\label{KKQT4}
\end{align}
\end{subequations}
where
\begin{align}
  \mathrm{G}_\pm(z,\mu) &= \int\limits_0^\infty d\omega \, e^{-i\omega z}g_\pm(\omega,\mu)
\qquad
{\rm and}
\qquad
   \bar\Lambda = m_B -m_b\,.
\end{align}
With the above relations, one can calculate the LCDAS $G_\pm$ with the leading twist and three-particle higher twist LCDAs.

\subsection{The phenomenological models}
Different from the LCDAs of light mesons which can be expanded in terms of Gagenbauer polynomials and the corresponding Gagenbauer moments can be calculated by Lattice or QCD sum rules since they are determined by the matrix elements of local operators, the LCDAs of $B$-meson is more difficult to be modelled. The evolution of $B$-meson can provide some model independent constraints to the behavior of the leading power LCDA of $B$-meson. In the large $\omega$ region, the operator product expansion (OPE) can be employed to explore the model-independent properties of the LCDA. The method is to calculate the first several moments of the distribution amplitude, derive its asymptotic behavior, and study its properties under renormalization-group evolution, then the constraints on the LCDAs can be found. The result indicates that at large $\omega$ region the LCDA falls off faster than $1/\omega$. In the low $\omega$ region the behavior of the LCDA cannot be constrained by the perturbative QCD, and only can be modelled with nonperturbative method. In practice, the $B$-meson LCDA is usually applied to the decay processes on $B$-meson, therefore, the spectator is regarded as a soft quark, and low $\omega$ region behavior is more important. Hereafter, we introduce some commonly  used models
The asymptotic behavior of those LCDAs at the small quark and glue momentum is relative to the conformal spins of the quark and glue field satisfied 
			\begin{equation}
\begin{aligned}
					\phi(\omega_1,\omega_2) \sim \omega_1^{2j_1 - 1} \omega_2^{2j_2-1}, \quad \phi \in \left \lbrace \phi_3, \phi_4, \psi_4 \cdots \right \rbrace.	
				  \label{asymptotic-behavior}
			\end{aligned}  \end{equation}
This relation can be obtained from the correlation function of the light-ray operators and suitable local current \cite{Braun:2017liq}.
Several two-particle and three-particle LCDAs to the twist-four accuracy have been given with a more general ansatz in \cite{Beneke:2018wjp}, such as the exponential model, the free parton model and local duality model, etc.. Similarly, we can obtain all the LCDAs models in accord with the correct low-momentum behaviour (\cite{Braun:2017liq}) and EOM constrains (tree level):
			\begin{eqnarray}
					\phi_{B_q}^+(\omega) &=& \omega f(\omega), \nonumber \\
					\phi_{B_q}^-(\omega) &=& \int_\omega^\infty d \rho \, f(\rho) + {1 \over 6}\varkappa (\lambda_E^2 - \lambda_H^2) \left [ \omega^2 f'(\omega) + 4\omega  f(\omega) - 2\int_\omega^\infty d \rho \, f(\rho) \right ], \nonumber \\
					g_-(\omega) &=& {1 \over 4} \int_\omega^\infty d x \, \big \{ (x - \omega) \left [ \phi_{B_q}^+(\omega) - \phi_{B_q}^-(\omega) \right ] - 2( \bar \Lambda_q - x) \phi_{B_q}^-(\omega) \big \} \nonumber \\
					&~&- {1 \over 2} \int_0^\omega d \omega_1 \, \int_{\omega - \omega_1}^\infty d \omega_2  \, {1 \over \omega_2} (1 - {\omega - \omega_1 \over \omega_2}) \psi_5(\omega_1, \omega_2), \nonumber \\
					\phi_3(\omega_1,\omega_2) &=& -{1 \over 2} \varkappa (\lambda_E^2 - \lambda_H^2)\omega_1 \omega_2^2 f'(\omega_1 + \omega_2), \nonumber \\
					\phi_4(\omega_1, \omega_2) &=& {1 \over 2} \varkappa (\lambda_E^2 + \lambda_H^2)\omega_2^2 f(\omega_1 + \omega_2), \nonumber \\
					\psi_4(\omega_1, \omega_2) &=& \varkappa \lambda_E^2 \omega_1 \omega_2 f(\omega_1 + \omega_2), \nonumber \\
					\phi_5(\omega_1, \omega_2) &=& \varkappa (\lambda_E^2 + \lambda_H^2)\omega_1 \int_{\omega_1 + \omega_2}^\infty d \omega \, f(\omega), \nonumber \\
					\psi_5(\omega_1,\omega_2) &=& \varkappa \lambda_E^2 \omega_2 \int_{\omega_1 + \omega_2}^\infty d \omega \, f(\omega), \nonumber \\
					\tilde \psi_5(\omega_1, \omega_2) &=& \varkappa \lambda_H^2 \omega_2 \int_{\omega_1 + \omega_2}^\infty d \omega \, f(\omega), \nonumber \\
					\phi_6(\omega_1, \omega_2) &=& \varkappa (\lambda_E^2 - \lambda_H^2) \int_{\omega_1 + \omega_2}^\infty d \omega \int_\omega^\infty d \omega ' \, f(\omega '),
			\end{eqnarray}
where the normalization constant
			\begin{equation}
\begin{aligned}
					\varkappa^{-1}={1\over 2}\int_0^\infty \omega^3 f(\omega) \, d \omega.
			\end{aligned}  \end{equation}
It is clear that this model is in agreement with the local duality model and the exponential model in \cite{Lu:2018cfc}. The parameter $\lambda_E$ and $\lambda_H$ in HQET are defined as the hadronic matrix element for both light and heavy quark with EOM constraints
\begin{equation} \begin{aligned}
					\langle 0 | \bar q \, g_s G_{\mu \nu} \Gamma \, h_v | \bar B \rangle = -{\tilde f_{B_q}m_{B_q} \over 6} \mbox{Tr}\Big\{ {1 + \slashed v \over 2} \left [ i\sigma_{\mu \nu} \lambda_H^2 + (v_\mu \gamma_\nu - v_\nu \gamma_\mu) (\lambda_H^2 - \lambda_E^2) \right ] \gamma_5 \Gamma \Big \}.
\end{aligned}  \end{equation}
The renormalization scale dependence of $\lambda_E$ and $\lambda_H$ can be obtain by solving the renormalization-group equations at the one-loop order
\begin{equation}
\left ( \begin{aligned} \lambda_E^2 (\mu) \\ \lambda_H^2 (\mu) \end{aligned} \right ) = \hat V \left [ \begin{aligned} \left ( {\alpha_s(\mu) \over \alpha_s(\mu_0)} \right )^{\gamma^{(0)}_i / (2 \beta_0)} \end{aligned} \right ]_{diag} \hat V^{-1} \left ( \begin{aligned} \lambda_E^2 (\mu_0) \\ \lambda_H^2 (\mu_0) \end{aligned} \right ).
\end{equation}
Here, we employ the three-parameter model of $\phi_+(\omega,\mu_0)=\omega f_{3p}(\omega,\mu_{0})$ where
\begin{align}
f_{3p}(\omega,\mu_{0}) = {\Gamma(\beta) \over \Gamma(\alpha)} {1 \over \omega_0^2} e^{-\omega / \omega_0} \text{U} \left ( \beta - \alpha, 3 - \alpha, \omega / \omega_0 \right )\,.
\end{align}
The three parameters $\alpha,\beta$ and $\omega_0$ determein the model of $B$-meson LCDAs at the initial scale $\mu=1.0$ GeV. The values of the logarithmic moments $\lambda_{B_q}$, $\hat \sigma_1$ and $\hat \sigma_2$ are needed, which is defined as \cite{Beneke:2018wjp}
		\begin{equation}
\begin{aligned}
				\hat{\sigma}_n(\mu)=\int_0^\infty d\omega \, {\lambda_B \over \omega}\ln^n {\lambda_Be^{-\gamma_E} \over \omega}\phi_{B_q}^+(\omega,\,\mu),
		\end{aligned}
		\label{shape-parameter}
		 \end{equation}
where $\lambda_{B_q}$ is defined by $\sigma_0 = 1$. In the terms of the three parameters $\omega_0, \alpha, \beta$ we obtain with a short calculation
		\begin{equation}
\begin{aligned}
				&\lambda_{B_q} = {\alpha - 1 \over \beta - 1} \omega_0, \qquad
				 \hat \sigma_1 = \psi(\beta-1)-\psi(\alpha-1)+\ln \left (\frac{\alpha-1}{\beta-1} \right ),\\
				&\hat \sigma_2 = {\pi^2 \over 6} + \left [ \psi(\beta-1)-\psi(\alpha-1)+\ln \left (\frac{\alpha-1}{\beta-1} \right ) \right ]^2 - \left [ \psi'(\beta-1)-\psi'(\alpha-1) \right ].
		\end{aligned}  \end{equation}

\section{LCSR with $B$-meson LCDA}
\label{sec:lcsr}

The LCSR with  $B$-meson LCDAs is widely employed to calculated the heavy-to-light form factors which receive the well-known end-point divergence.
Starting from the QCD two-point correlation functions at Euclidean space, the $B$-meson LCSR can avoid the end-point divergence with the help of the dispersion relation, the Borel transformation, the quark-hadron duality.
The heavy-to-light form factors can also be calculated from the SCET correlation function which will be called the SCET improved LCSR hereafter.
Below we will calculate the $B\to P$ form factors and the $B\to V$ form factors with the $B$-meson LCSR and the SCET improved LCSR, respectively.

\subsection{The heavy-to-light form factors}
The form factors for $\bar{B}$ decays into a pseudoscalar meson are
defined by the matrix elements of the weak transition current sandwiched between the $B$-meson and a pseudocalar meson states, i.e.
\begin{align}
\langle P(p)|\bar q \, \gamma^\mu b |\bar{B}(p_{B})\rangle =&~
f_{B\to P}^+(q^2)\left[p_{B}^\mu+p^{\mu}-\frac{m_{B}^2-m_P^2}{q^2}\,q^\mu\right]
+f_{B\to P}^0(q^2)\,\frac{m_{B}^2-m_P^2}{q^2}\,q^\mu,
\nonumber \\
\langle P(p)|\bar q \, \sigma^{\mu\nu} q_\nu b|\bar{B}(p_{B}) \rangle =&~
\frac{i f_{B\to P}^T(q^2)}{m_{B}+m_P}\left[q^2(p_{B}^\mu+p^{\mu})-
(m_{B}^2-m_P^2)\,q^\mu\right],
\label{ftensor}
\end{align}
where $m_{B}$ and $m_{P}$ are respectively the masses of the $B$ meson and the pseudoscalar
meson and the momentum transfer $q=p_{B}-p$. The relevant form factors for $B$ decays into vector
mesons are defined as
\begin{align}
\langle V(p,\varepsilon^\ast)| \bar q \gamma^\mu b | \bar{B}(p_{B}) \rangle =&~ 
 \frac{2iV(q^2)}{m_{B}+m_V} \,\epsilon^{\mu\nu\rho\sigma}
 \varepsilon^{\ast}_\nu \, p_\rho p_{B,\sigma},
  \nonumber \\
\langle V(p,\varepsilon^\ast)| \bar q \gamma^\mu\gamma_5 b | \bar{B}(p_{B})
\rangle =&~
  2m_VA_0(q^2)\,\frac{\varepsilon^\ast\cdot q}{q^2}\,q^\mu +
  (m_{B}+m_V)\,A_1(q^2)\left[\varepsilon^{\ast\mu}-
  \frac{\varepsilon^\ast\cdot q}{q^2}\,q^\mu\right]
\nonumber\\[0.0cm]
  & 
-\,A_2(q^2)\,\frac{\varepsilon^\ast\cdot q}{m_{B}+m_V}
 \left[p_{B}^\mu+p^{\mu} -\frac{m_{B}^2-m_V^2}{q^2}\,q^\mu\right],
 \nonumber \\
\langle V(p,\varepsilon^\ast)| \bar q \sigma^{\mu\nu}q_\nu b | \bar{B}(p_{B})
\rangle =&~
  2\,T_1(q^2)\,\epsilon^{\mu\nu\rho\sigma}\varepsilon^{\ast}_\nu\,
  p_{B,\rho} p_\sigma,
\nonumber \\
\langle V(p,\varepsilon^\ast)| \bar q \sigma^{\mu\nu} \gamma_5 q_\nu b |
\bar{B}(p_{B}) \rangle=&~
(-i)\,T_2(q^2)\left[(m_{B}^2-m_V^2)\,\varepsilon^{\ast\mu}-(\varepsilon^\ast\cdot
q)\,(p_{B}^\mu+p^{\mu})\right]
\nonumber\\[0.0cm]
 &
+\,(-i)\,T_3(q^2)\,(\varepsilon^\ast\cdot
q)\left[q^\mu-\frac{q^2}{m_{B}^2-m_V^2}(p_{B}^\mu+p^{\mu})\right],
\label{ffdef}
\end{align}
where $m_V$ ($\varepsilon$) is the mass (polarisation vector)
of the vector meson and we use the sign convention $\epsilon^{0123}=-1$.
At maximal hadronic recoil $q^2=0$ there exist  three  relations for
the above-mentioned $B \to M$ form factors in QCD
\begin{eqnarray}
f_{B\to P}^+(0)=f_{B\to P}^0(0),\qquad {m_B + m_V \over 2 \, m_V} \, A_1(0) - {m_B - m_V \over 2 \, m_V} \, A_2(0) = A_0(0) \,,
\qquad T_1(0) = T_2(0) \,,
\end{eqnarray}
which are free of both radiative and power corrections.

\subsection{The $B \to P$ form factors with LCSR}
In this subsection we aim at obtain the light-cone sum rules for the $B \to P$ form factors at the leading power on $1/m_b$ and leading order on $\alpha_s$. In order to obtain the light-cone sum rules for $B \to P$ form factors, we use the following vacuum-to-$B$-meson correlation function:
\begin{equation}
	\begin{aligned}
		\Pi_{\mu} (n \cdot p, \bar n \cdot p) & = \int d^4 x \, e^{i p \cdot x} \langle 0 | T \Big \{
		\bar d(x)\, \slashed n \gamma_5 \,q(x),\, \bar q(0)\, \Gamma_\mu \,b(0) \Big \} | B(p_{B}) \rangle
		\\
		& =  \left\{
		\begin{array}{l}
			\Pi(n \cdot p, \bar n \cdot p)\, n_\mu + \widetilde \Pi (n \cdot p, \bar n \cdot p)\, \bar n_\mu, \hspace{1cm} \Gamma_\mu = \gamma_\mu \vspace{0.5cm} \\
			\Pi_T(n \cdot p, \bar n \cdot p)\, \left [ \bar n \cdot q \, n_\mu - n \cdot q \, \bar n_\mu \right ], \hspace{1.25cm} \Gamma_\mu = \sigma_{\mu \nu} q^\nu
\end{array},
 \hspace{0.5 cm} \right.
 \label{correlation_function}
\end{aligned}  \end{equation}
where the Lorentz structure $\Gamma_\mu$ stands for the two different $b\to q$ weak currents in QCD. 
In the center-of-mass frame, the $B$-meson momentum $p_{B} = p + q = m_{B} v$, where $p,\, q$ stand for the momentum of the light-meson and the weak current. At large recoil $n \cdot p \sim m_b$, we work with $p^2<0$ in order that the correlation function can be calculated perturbatively. We take $\bar n$ standing for the direction of $p$ and use the following power counting with the representation of $q \sim (n \cdot q,\bar n \cdot q, q_\perp)$ :
\begin{equation}
\begin{aligned}
p_{B} \sim m_b \,(1, 1, 1), 
\qquad
p \sim m_b \,(1, \lambda^{2}, \lambda), 
\qquad
\qquad
k \sim m_b \,(\lambda^2, \lambda^2, \lambda^2),
\end{aligned}  \end{equation}
where the power counting parameter $\lambda = \sqrt{{\Lambda_{\rm{QCD}} / m_b}}$, $k$ is the momentum of the light-quark in $B$-meson, and $m_b$ is b-quark mass. 
For the interpolation current, we employ $\bar d(x)\, \slashed n \gamma_5 \,q(x)$ so that we can obtain the leading power contribution to the form factors directly. 

To guarantee the validity of the light-cone operator-product expansion (OPE), one must prove the light-cone dominance of the correlation functions. In the correlation functions, we have assumed that both four-momenta are  spacelike, $p^2<0$, and  sufficiently large: $P^2 =-p^2\gg \Lambda_{\rm QCD}^2$, in addition, the ratio $\xi={2 p\cdot k }/{P^2}\sim {\cal O}(1)$, the  $x^2\sim 1/P^2$ must be satisfied under the condition that the exponent  $e^{ipx}$ does not oscillate strongly. This leads to a constraint on region of the $q^2$ which is accessible to OPE on the light-cone in the $B\to P$ transitions
\begin{eqnarray}
0\leq q^2 <  m_b^2 -m_b P^2/\Lambda_{\rm QCD}.
\label{eq-interv}
\end{eqnarray}
The light-cone dominance of the
correlation function allows one to
contract the $q$ and $\bar{q}$ fields and use the free-quark propagator
$S_{q}(x)=-i\langle 0 | q(x)\bar{q}(0)|0\rangle$
as a leading-order approximation.
Then the correlation function can be written by:
\begin{eqnarray}
\Pi_{\mu} (n \cdot p, \bar n \cdot p) = \int d^4 x \, e^{i p \cdot x} [\slashed n \gamma_5S_{q}(x)\Gamma_\mu ]_{\alpha\beta}\langle 0 |
		\bar d(x)_\alpha\,   \,h_v(0)_\beta\,  | B(p_{B}) \rangle
\end{eqnarray}
where the heavy quark field have been expanded using HQET and only the large component has been kept.  If one expands the bilocal operator
$\bar{q}_{2\alpha}(x) h_{v\beta}(0)$ in the small $x$ region,
an infinite  series of matrix elements of local operators are required for the
vacuum-pion amplitudes. Instead, one has to retain in
the matrix element of the operator, which introduces the LCDAs of $B$-meson. The detailed discussion on the LCDAs of $B$-meson has been done in the previous section.

We will use the calculation of the form factors $f^{+}_{B\to P}$ and $f^{0}_{B\to P}$ as an illustration of the LCSR approach.
With $\Gamma_\mu = \gamma_\mu$, the correlation function is then expressed in terms of the convolution of the hard scattering kernel and the LCDAs of $B$ meson, and at tree level the result reads
\begin{eqnarray}
\widetilde{\Pi}(n \cdot p,\bar n \cdot p) &=& \tilde f_B(\mu) \, m_B \,
\int_0^{\infty} d \omega \,
\frac{\phi_-(\omega)}{\omega - \bar n \cdot p- i \, 0}
+ {\cal O}(\alpha_s) \,, \nonumber \\
\Pi(n \cdot p,\bar n \cdot p) &=& {\cal O}(\alpha_s) \,.
\label{factorization of correlator:tree}
\end{eqnarray}
In order to arrive at the sum rules, one has to express this result in terms of the dispersion integral with respect to $\bar n\cdot p$
\begin{align}
\widetilde{\Pi}_{\rm QCD}(n \cdot p,\bar n \cdot p) 
=& \int^{\infty}_{0} \frac{d\omega'}{\omega'-\bar n\cdot p}
{\rm Im_{\omega'}} \widetilde{\Pi}(n \cdot p,\omega')\,,
\end{align}
and ${\Pi}(n \cdot p,\omega')$ should also be expressed in the dispersion form.
The relative hadronic representation of the vacuum-to-$B$-meson correlation function and the decay constant of the pseudoscalar meson are given by
\begin{align}
\Pi^{had}_{\mu}(P_{B}, p) =&~ {\langle P(p) | \bar d\, \slashed n \gamma_5 \,q | 0 \rangle \langle P(p) | \bar q\, \gamma_\mu \,b | B_q(P_{B}) \rangle \over m_P^2 - p^2} + {\rm continuum}, \\
=&~ \frac{f_{P} \, m_B}{2 \, (m_{P}^2/ n \cdot p - \bar n \cdot p)}
\bigg \{  \bar n_{\mu} \, \left [ \frac{n \cdot p}{m_B} \, f_{B \to P}^{+} (q^2) + f_{B \to P}^{0} (q^2)  \right ]
\nonumber \\
& +  \,  n_{\mu} \, \frac{m_B}{n \cdot p-m_B}  \, \,
\left [ \frac{n \cdot p}{m_B} \, f_{B \to P}^{+} (q^2) -  f_{B \to P}^{0} (q^2)  \right ] \bigg \} \, \nonumber \\
&+ \int_{\omega^{P}_s}^{\infty}   \, \frac{d \omega^{\prime} }{\omega^{\prime} - \bar n \cdot p - i 0} \,
\left [ \rho_{n}^{had}(n \cdot p,\omega^{\prime})  \, n_{\mu} \,
+\rho_{\bar n}^{had}(n \cdot p,\omega^{\prime})  \, \bar{n}_{\mu}  \right ] \,,
	 \label{hadronic}
\end{align}
where we have used the definitions of the form factors (\ref{ftensor}) and the light-meson decay constant
\begin{align}
\langle 0 |  \bar d\, \slashed n \gamma_5 \,q |P(p) \rangle = i n \cdot p f_P \,.
\end{align}
The parameter $\omega_s^P = s^P_0 / n \cdot p \sim \lambda^2$ corresponds to the hadronic threshold of the light-meson channel.
Taking advantage of the parton-hadron duality assumption, one can relate the result from the parton level calculation to the parametrization in the hadronic level  
\begin{align}
\int_{\omega^{P}_s}^{\infty}   \, \frac{d \omega^{\prime} }{\omega^{\prime} - \bar n \cdot p} \,
\rho_{\bar n}^{had}(n \cdot p,\omega^{\prime})
=&~ \int_{\omega^{P}_s}^{\infty}   \, \frac{d \omega^{\prime} }{\omega^{\prime} - \bar n \cdot p} \,
{\rm Im_{\omega'}} \widetilde{\Pi}(n \cdot p,\omega')\,,
\end{align}
a similar relation also hold for $\rho_{n}^{had}(n \cdot p,\omega^{\prime})$ and ${\Pi}(n \cdot p,\omega^{\prime})$.
Then we should perform the Borel transformation with respect to the variable $\bar n \cdot p $ to both the hadronic and partonic representation of the correlation \cite{Colangelo:2000dp}
\begin{align}
\Pi_{\cal B}(\omega_{M})\equiv 
{\cal B}_{\omega_{M}} \Pi(\bar n\cdot p) = \mathop{{\rm lim}}\limits_{-\bar n\cdot p, r\to \infty}^{-\bar n\cdot p/r=\omega_{M}} \frac{(-\bar n\cdot p)^{r+1}}{r!}
\left(\frac{d}{d\bar n\cdot p}\right)^{r} \Pi(\bar n\cdot p)\,.
\end{align} 
With the above procedures, one can obtain the sum rules for  the form factors
\begin{eqnarray}
f_{B\to P}^{+}(q^2) &=& \frac{\tilde f_B(\mu) \, m_B}{f_{\pi} \,n \cdot p} \, {\rm exp} \left[{m_{P}^2 \over n \cdot p \,\, \omega^{P}_M} \right]
\int_0^{\omega^{P}_s} \, d \omega^{\prime} \, e^{-\omega^{\prime} / \omega^{P}_M} \,  \phi_B^{-}(\omega^{\prime}) + {\cal O}(\alpha_s) \,,
\nonumber \\
f_{B\to P}^{0} (q^2) &=&  \frac{n \cdot p}{m_B} \, f_{B\to P}^{+} (q^2) +  {\cal O}(\alpha_s) \,.
\label{the form-factor relation}
\end{eqnarray}
where the Borel mass $ \omega^{P}_M = M^2_{P} / n \cdot p \sim \lambda^2$. At tree level as well as leading power, the scalar  correlation functions $\Pi(q^2)$ vanishes, which exhibits the large recoil symmetry of the form factors. The tensor form factor can also be calculated in the same method by using the correlation function with $\Gamma_\mu = \sigma_{\mu \nu} q^\nu$.

\subsection{The $B \to V$ form factors SCET improved LCSR}

In this subsection we will calculate the $B \to V$ form factors with the SCET improved LCSR.
First we match QCD onto ${\rm SCET}$ where the
heavy-to-light form factors are given by \cite{Beneke:2003pa,Lange:2003pk,Bauer:2002aj}
\begin{eqnarray}
(\bar \psi \, \Gamma_i \, Q)(0) &=& \int d {\hat s} \, \sum_{j} \, \tilde{C}_{i j}^{(\rm A0)}(\hat s) \,
O_{j}^{(\rm A0)}(s; 0) + \int d  {\hat s} \, \sum_{j} \, \tilde{C}_{i j \mu}^{(\rm A1)}(\hat s) \,
O_{j}^{(\rm A1) \mu}(s; 0) \nonumber \\
&& +  \int d  {\hat s_1} \, \int d  {\hat s_2} \, \sum_{j} \, \tilde{C}_{i j \mu}^{(\rm B1)}(\hat s_1, \hat s_2) \,
O_{j}^{(\rm B1) \mu}(s_1, s_2; 0) + ...    \,,
\end{eqnarray}
The hard matching coefficients for both the ${\rm A0}$-type and ${\rm B1}$-type SCET currents have been
computed at one-loop accuracy \cite{Hill:2004if,Bauer:2000yr,Beneke:2004rc}.
The seven QCD $B \to V$ form factors are expressed in terms of the four ``effective"
form factors in ${\rm SCET}$ at leading power in the heavy quark expansion \cite{Beneke:2005gs}
\begin{eqnarray}
&& \langle V(p, \epsilon^{\ast}) | \left (\bar \xi \, W_c \right) \,
\gamma_5 \, h_v \,   | \bar B_v \rangle  =
- n \cdot p \, (\epsilon^{\ast} \cdot v) \, \xi_{\parallel}(n \cdot p) \,, \nonumber \\
&& \langle V(p, \epsilon^{\ast}) | \left (\bar \xi \, W_c \right) \,
\gamma_5 \, \gamma_{\mu \perp} \, h_v \,   | \bar B_v \rangle  =
- n \cdot p \, (\epsilon^{\ast}_{\mu} - \epsilon^{\ast} \cdot v \, \bar n_{\mu}) \,
\xi_{\perp}(n \cdot p) \,, \nonumber \\
&& \langle V(p, \epsilon^{\ast}) | \left (\bar \xi \, W_c \right) \,
\gamma_5 \, \left (W_c^{\dagger} \,\, i \, \not \! \! D_{c \perp} \, W_c \right)(r n) \, h_v \,   | \bar B_v \rangle
=  - n \cdot p  \, m_b \, \epsilon^{\ast} \cdot v \,
\int_0^1 \, d \tau \, e^{i \, \tau \, n \cdot p \, r} \, \Xi_{\parallel}(\tau, \, n \cdot p)  \,, \nonumber \\
&& \langle V(p, \epsilon^{\ast}) | \left (\bar \xi \, W_c \right) \,
\gamma_5 \, \gamma_{\mu \perp} \,
\left (W_c^{\dagger} \,\, i \, \not \! \! D_{c \perp} \, W_c \right)(r n) \, h_v \,   | \bar B_v \rangle  \nonumber \\
&& =  - n \cdot p  \, m_b \,  (\epsilon^{\ast}_{\mu} - \epsilon^{\ast} \cdot v \, \bar n_{\mu}) \,
\int_0^1 \, d \tau \, e^{i \, \tau \, n \cdot p \, r} \, \Xi_{\perp}(\tau, \, n \cdot p)  \,,
\label{Definition: SCET-I form factors}
\end{eqnarray}
where  the light-cone Wilson line is introduced to restore the collinear gauge invariance \cite{Beneke:2003pa,Beneke:2002ni}
\begin{eqnarray}
W_c(x) =  {\rm P} \, {\rm exp}  \,
\left [ i \, g_s \, \int_{-\infty}^{0} \, d s \, n \cdot A_c(x + s \, n) \right ] \,.
\label{collinear Wilson line}
\end{eqnarray}
The relation between the QCD form factors and the SCET form factors are \cite{Beneke:2005gs}
\begin{eqnarray}
f^{i}_{B \to V}(n \cdot  p) = C_i^{\rm (A0)}(n \cdot  p) \, \xi_a(n \cdot p)
+ \int d \tau \, C_i^{\rm (B1)}(\tau, n \cdot p) \, \Xi_a(\tau, n \cdot p) \,, \,\,\,
(a \, = \, \parallel,  \,\,  \perp)\,,
\label{SCET-I factorization formula for B to V FFs}
\end{eqnarray}
where
\begin{eqnarray}
&& {m_B \over m_B + m_V} \, V(n \cdot p)  =  C_V^{(\rm A0)} \, \left ({n \cdot p \over m_b}, \mu \right ) \,
\xi_{\perp}(n \cdot p)  \nonumber \\
&& \hspace{4 cm} + \int_0^1  d \tau \, C_V^{(\rm B1)} \,
\left ({n \cdot p \, \bar \tau \over m_b},
 {n \cdot p \, \tau \over m_b},\mu \right )  \, \Xi_{\perp}(\tau, n \cdot p) \,,  \nonumber \\
&& {2 \, m_V \over n \cdot p} \, A_0(n \cdot p)  =  C_{f_0}^{(\rm A0)} \, \left ({n \cdot p \over m_b}, \mu \right ) \,
\xi_{\parallel}(n \cdot p)   + \int_0^1  d \tau \, C_{f_0}^{(\rm B1)} \,
\left ({n \cdot p \, \bar \tau \over m_b},
 {n \cdot p \, \tau \over m_b},\mu \right )  \, \Xi_{\parallel}(\tau, n \cdot p) \,,  \nonumber  \\
&& {m_B + m_V \over n \cdot p} \, A_1(n \cdot p)  =  C_V^{(\rm A0)} \, \left ({n \cdot p \over m_b}, \mu \right ) \,
\xi_{\perp}(n \cdot p) \nonumber \\
&& \hspace{4 cm} + \int_0^1  d \tau \, C_V^{(\rm B1)} \,
\left ({n \cdot p \, \bar \tau \over m_b},
 {n \cdot p \, \tau \over m_b},\mu \right )  \, \Xi_{\perp}(\tau, n \cdot p) \,,  \nonumber  \\
&& {m_B + m_V \over n \cdot p} \, A_1(n \cdot p)
- {m_B - m_V \over m_B} \, A_2(n \cdot p) \nonumber \\
&& =  C_{f_+}^{(\rm A0)} \, \left ({n \cdot p \over m_b}, \mu \right ) \,
\xi_{\parallel}(n \cdot p)   + \int_0^1  d \tau \, C_{f_+}^{(\rm B1)} \,
\left ({n \cdot p \, \bar \tau \over m_b},
 {n \cdot p \, \tau \over m_b},\mu \right )  \, \Xi_{\parallel}(\tau, n \cdot p) \,,  \nonumber  \\
&& T_1(n \cdot p)  =  C_{T_1}^{(\rm A0)} \, \left ({n \cdot p \over m_b}, \mu \right ) \,
\xi_{\perp}(n \cdot p)  + \int_0^1  d \tau \, C_{T_1}^{(\rm B1)} \,
\left ({n \cdot p \, \bar \tau \over m_b},
 {n \cdot p \, \tau \over m_b},\mu \right )  \, \Xi_{\perp}(\tau, n \cdot p) \,,  \nonumber \\
&& {m_B \over n \cdot p} \,T_2(n \cdot p)  =  C_{T_1}^{(\rm A0)} \, \left ({n \cdot p \over m_b}, \mu \right ) \,
\xi_{\perp}(n \cdot p)  + \int_0^1  d \tau \, C_{T_1}^{(\rm B1)} \,
\left ({n \cdot p \, \bar \tau \over m_b},
 {n \cdot p \, \tau \over m_b},\mu \right )  \, \Xi_{\perp}(\tau, n \cdot p) \,,  \nonumber \\
&& {m_B \over n \cdot p} \,T_2(n \cdot p) - T_3(n \cdot p) \nonumber \\
&& =  C_{f_T}^{(\rm A0)} \, \left ({n \cdot p \over m_b}, \mu \right ) \,
\xi_{\parallel}(n \cdot p)  + \int_0^1  d \tau \, C_{f_T}^{(\rm B1)} \,
\left ({n \cdot p \, \bar \tau \over m_b},
 {n \cdot p \, \tau \over m_b},\mu \right )  \, \Xi_{\parallel}(\tau, n \cdot p) \,.
\label{SCET-I factorization formulae}
\end{eqnarray}
The coefficient functions $C_{i j}^{(\rm A0)}$ and $C_{i j \mu}^{(\rm B1)}$ are obtained from the
Fourier transformations of the position-space coefficient functions
$\tilde{C}_{i j}^{(\rm A0)}$ and $\tilde{C}_{i j \mu}^{(\rm B1)}$ \cite{Beneke:2003pa}.
It is evident that only five independent combinations of ${\rm A0}$- and ${\rm B1}$-type  SCET operators
appear in the factorization formulae for the seven different $B \to V$ form factors,
implying the two additional relations \cite{Beneke:2000wa,Burdman:2000ku}
\begin{eqnarray}
{m_B \over m_B + m_V} \, V(n \cdot p)  = {m_B + m_V \over n \cdot p} \, A_1(n \cdot p) \,,
\qquad  T_1(n \cdot p) =  {m_B \over n \cdot p} \,T_2(n \cdot p) \,,
\label{exact form factor relations at LP}
\end{eqnarray}
which are fulfilled to all orders in perturbative expansion at leading power in $\Lambda/m_b$.

With the relation between of QCD form factors at hand, we only need to calculate the SCET form factors $\xi_i$ and $\Xi_i$.
Within the framework of light-cone sum rules, one can employ either the weak current in full QCD and the weak current in the SCET in the definition of the correlation function, the former case will lead to the from factors in the QCD and the latter will lead to the SCET improved LCSR for the form factors $\xi_i$ and $\Xi_i$. 

We will use the calculation of the SCET form factor $\xi_{\parallel}(n \cdot p) $ as an example to illustrate the application of SCET improved LCSR.
In the SCET sum rules, the correlation function is constructed with the field in the SCET. 
We start with the the vacuum-to-$B$-meson correlation function
\begin{eqnarray}
\Pi_{\nu, \|}(p, q) = \int d^4 x \, e^{i p \cdot x} \, \langle 0 |
{\rm T}  \left \{j_{\nu}(x),  \,\, \left (\bar \xi \, W_c \right)(0) \,
\gamma_5 \, h_v(0) \,    \right \}   | \bar B_v \rangle \,,
\label{correlation function for xi-L}
\end{eqnarray}
where the local QCD current $j_{\nu}$ interpolates current for the longitudinal polarization state
of the collinear vector meson \cite{Beneke:2002ph}
\begin{eqnarray}
j_{\nu}(x) = \bar q^{\prime}(x) \, \gamma_{\nu} \, q(x)
= j_{\xi \xi, \nu}^{(0)} + j_{\xi \xi, \perp \, \nu}^{(1)} \,
+ j_{\xi q_s, \parallel \, \nu}^{(2)}  + j_{\xi q_s, \perp \, \nu}^{(2)} + ... \,, \,,
\end{eqnarray}
where the explicit expressions of the effective currents are given by
\begin{eqnarray}
j_{\xi \xi, \nu}^{(0)} &=& \bar \xi \, {\not \! n \over 2} \, \xi \,\, \bar n_{\nu} \,,  \nonumber \\
j_{\xi \xi, \perp \,  \nu}^{(1)}  &=&   \bar \xi \, \gamma_{\nu \perp} \,
{1 \over i \, n \cdot D_c} \, i \not \! \! D_{c \perp}\,  {\not \! n \over 2} \, \xi
+ \bar \xi \, i \not \! \! D_{c \perp}\,
{1 \over i \, n \cdot D_c} \, \gamma_{\nu \perp} \,  {\not \! n \over 2} \, \xi   \,,  \nonumber \\
j_{\xi q_s, \parallel \, \nu}^{(2)} &=& \left ( \bar \xi \, W_c \,  {\not \! n \over 2} \, Y_s^{\dagger} \, q_s
+ \bar q_s \, Y_s \,  {\not \! n \over 2} \,  W_c^{\dagger} \, \xi \right ) \,\, \bar n_{\nu} \,, \nonumber \\
j_{\xi q_s, \perp \, \nu}^{(2)} &=&  \bar \xi \, W_c \,   \gamma_{\perp \nu}  \, Y_s^{\dagger}\, q_s
+ \bar q_s   \,   Y_s \,  \gamma_{\perp \nu} \, W_c^{\dagger} \, \xi \,,
\end{eqnarray}
where the collinear Wilson line
defined in (\ref{collinear Wilson line}) and the following soft Wilson line
\begin{eqnarray}
Y_s(x) &=&  {\rm P \,\, exp} \left [i \, g_s \, \int_{-\infty}^{0} \, ds \,  \bar n \cdot A_s(x + s \bar n) \right ] \,,
\end{eqnarray}
is introduced to keep gauge invariance. It is then straightforward to write down the leading-power contribution to the correlation function at tree level
\begin{eqnarray}
&& \Pi_{\nu, \|}(p, q) 
={\tilde{f}_B(\mu) \, m_B \over 2}\,
\int_0^{+\infty} \, d \omega \, {1 \over \bar n \cdot p - \omega+ i0}  \,
\phi_{-}(\omega, \mu) \,\, \bar n_{\nu}
\label{xi-LO}
\end{eqnarray}

Now we apply the standard method of sum rules, and match the spectral representation of the factorization formula (\ref{xi-LO})
with the corresponding hadronic dispersion relation
\begin{eqnarray}
&& \Pi^{had}_{\nu, \|}(p, q) \nonumber \\
&& =  \left [- {f_{V, \|}  \, m_V \over m_V^2/n \cdot p -\bar n \cdot p - i 0}  \,
\left ( {n \cdot p \over 2 \, m_V} \right )^2 \, \xi_{\|}(n \cdot p)
+ \int_{\omega^{V}_s}^{+\infty} \, {d \omega^{\prime} \over \omega^{\prime}  - \bar n \cdot p - i0} \,\,
\rho^{had}_{\|}(\omega^{\prime}, n \cdot p) \right ] \, \bar n_{\nu} \,. \hspace{0.8 cm}
\end{eqnarray}
to obtain the SCET form factor  $\xi_{\|}(n \cdot p)$
\begin{eqnarray}
&& \xi_{\|}(n \cdot p) = 2 \, {\tilde{f}_B(\mu)  \over f_{V, \|} } \,
{ m_B \, m_V \over (n \cdot p)^2} \, \int_{0}^{\omega^{V}_s} \, d \omega^{\prime} \,
{\rm exp} \left [ - {n \cdot p \, \omega^{\prime} - m_V^2 \over n \cdot p \, \omega^{V}_M} \right ] \phi_-(\omega^{\prime}, \mu)  \,.
\hspace{0.5 cm}
\end{eqnarray}
The scale-independent longitudinal decay constant of the vector meson is defined as follows
\begin{eqnarray}
c_V \, \langle V(p, \epsilon^{\ast}) | j_{\nu} | 0 \rangle = - i \, f_{V, \| } \,\,  m_V \,\,  \epsilon^{\ast}_{\nu}(p) \,.
\end{eqnarray}
The other SCET form factors $\xi_i$ and $\Xi_i$ can be calculated with the same procedure from different SCET correlation functions (\ref{correlation function for xi-L}).

\section{ QCD corrections to the $B\to M$ form factors}
\label{sec:nlo}

In this section we will show the calculation of the QCD corrections to the $B\to M$ form factors.
First we calculation the NLO correction to the $B\to P$ form factors with traditional LCSR with $B$-meson LCDAs.
Then we provide the NLO computation of the $B\to V$ form factors with SCET improved LCSR.

\subsection{The $B \to P$ form factors with $B$-meson LCSR}
The objective of this section is to establish the factorization formulae
 for $\Pi_{\mu}(n \cdot p,\bar n \cdot p)$  in QCD at one-loop level. In the previous section we have seen that the correlation function can factorized into the hard scattering part and the LCDAs of $B$-meson. Since the $B$-meson LCDA is nonperturbative objective, what we need is to find the QCD correction to the hard scattering part. The correlator  $\Pi_{\mu,  \, b \bar d}$ can be expanded as
\begin{eqnarray}
\Pi_{\mu,  }&=&\Pi_{\mu  \,}^{(0)}+ \Pi_{\mu \,}^{(1)} + ...
= \Phi_{b \bar q}  \otimes T \, \nonumber \\
&=&   \Phi_{b \bar q}^{(0)} \otimes  T^{(0)}
+  \left [ \Phi_{b \bar q}^{(0)} \otimes  T^{(1)}
+ \Phi_{b \bar q}^{(1)} \otimes  T^{(0)} \right ]  + ...  \,,
\end{eqnarray}
where $\otimes$ denotes the convolution in the variable $\omega^{\prime}$,
and $T^{(1)}$ is the one-loop level hard scattering kernel, it is then determined
by the matching condition
\begin{eqnarray}
\Phi_{b \bar q}^{(0)} \otimes  T^{(1)}
= \Pi_{\mu}^{(1)} -  \Phi_{b \bar q}^{(1)} \otimes  T^{(0)} \,,
\label{matching condition of T1}
\end{eqnarray}
where the second term serves as the infrared (soft) subtraction. We will demonstrate that the soft divergence will completely absorbed into the LCDA of $B$-meson and the there is no leading contribution to the correlation function from the collinear region (with the momentum scaling $l_{\mu} \sim(1, \lambda^2, \lambda)$) at leading power, which confirm the factorization at one-loop level. As a result, the  hard-scattering kernel $T$ can be  contributed only
from hard and/or hard-collinear regions at leading power in $\Lambda/m_b$. We will evaluate the master formula of $T^{(1)}$ in
 (\ref{matching condition of T1}) diagram by diagram.  There exist to typical perturbative  scale, namely, the hard scale $m_b$ and the hard-collinear scale $\sqrt{m_b\Lambda}$, therefore,  the hard scattering part can be further factorized into the hard function and the jet function. It is more convenient to apply the method of regions
\cite{Beneke:1997zp} to compute the loop integrals in order to obtain the hard coefficient
function ($C$) and the jet function ($J$) simultaneously.
$C$ and $J$ must be well defined in dimensional regularization.
This guarantees that we can adopt dimensional regularization to evaluate the leading-power
contributions of $\Pi_{\mu,  \, b \bar q}$ without introducing an additional ``analytical" regulator. In the following we will calculate the contribution in figure \ref{fig: NLO diagrams of the correlator}(a) in details and present the final results for the other diagram directly.
\begin{itemize}
\item

{\bf Weak vertex diagram}

The contribution to $\Pi_{\mu}^{(1)}$ from the QCD correction to the weak vertex
(figure \ref{fig: NLO diagrams of the correlator}(a)) is
\begin{eqnarray}
\Pi_{\mu,  \, weak}^{(1)}
&=& \frac{g_s^2 \, C_F}{2 \, (\bar n \cdot p -\omega)} \,
\int \frac{d^D \, l}{(2 \pi)^D} \,   \frac{1}{[(p-k+l)^2 + i 0][(m_b v+l)^2 -m_b^2+ i 0] [l^2+i0]}  \nonumber  \\
&& \bar d(k)  \! \not n \,\gamma_5  \,\! \not {\bar n} \,\, \gamma_{\rho}  \, (\! \not p - \! \not k  + \! \not l)
\, \gamma_{\mu} \, (m_b  \! \not v +  \! \not l+ m_b )\, \gamma^{\rho} \, b(v)  \,,
\label{diagram a: expression}
\end{eqnarray}
where   $D= 4 -2 \, \epsilon$. 
For  the following scalar integral
\begin{eqnarray}
I_1= \int [d \, l] \, \frac{1}{[(p-k+l)^2 + i 0][(m_b v+l)^2 -m_b^2+ i 0] [l^2+i0]} \,
\end{eqnarray}
the power counting of $I_1$ is $I_1 \sim \lambda^0$  for the three leading regions (the hard, hard-collinear and soft regions), thus only the leading-power contributions of the numerator in  (\ref{diagram a: expression}) need to be kept for a given region.
We define the integration measure as
\begin{eqnarray}
[d \, l] \equiv \frac{(4 \, \pi)^2}{i} \, \left ( \frac{\mu^2 \, e^{\gamma_E}}{4 \, \pi}\right )^{\epsilon} \,
\frac{d^D \, l}{(2 \pi)^D}\,.
\end{eqnarray}

\begin{figure}
\begin{center}
\includegraphics[width=.85 \columnwidth]{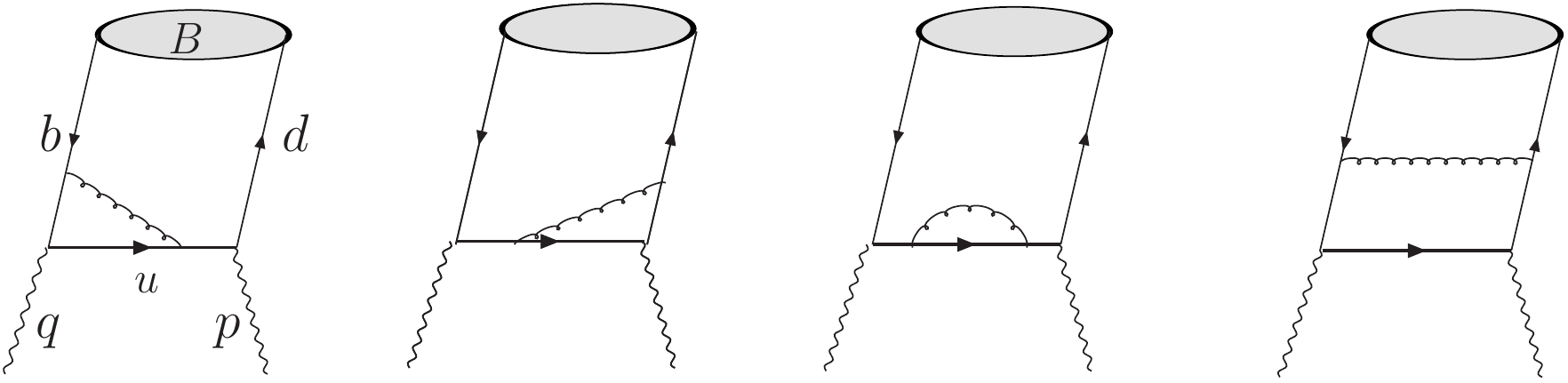}\\
\hspace{-0.5 cm}(a) \hspace{3.0 cm} (b)\hspace{3.5 cm} (c) \hspace{3.5 cm} (d) \\
\vspace*{0.1cm}
\caption{Diagrammatical representation of the correlation function
$\Pi_{\mu}(n \cdot p,\bar n \cdot p)$ at ${\cal O}(\alpha_s)$. }
\label{fig: NLO diagrams of the correlator}
\end{center}
\end{figure}

Inserting the partonic light-cone projector  yields the hard contribution
of $\Pi_{\mu, weak}^{(1)}$ at leading power
\begin{eqnarray}
\Pi_{\mu,  \, weak}^{(1), \, h}&=& i \, g_s^2 \, C_F \, \tilde f_B(\mu) \, m_B \, 
\frac{\phi_{b \bar q}^{-}(\omega)}{\bar n \cdot p -\omega} \int \frac{d^D \, l}{(2 \pi)^D} \, \nonumber \\
&&   \frac{1}{[l^2 + n \cdot p \,\, \bar n \cdot l+ i 0][l^2 + 2 \, m_b \, v \cdot l+ i 0] [l^2+i0]}  \nonumber  \\
&&  \times  \bigg \{ \bar n_{\mu}  \left [ 2 \, m_b \, n \cdot (p+l)  + (D-2)\, l_{\perp}^2 \right ]
- n_{\mu}  \, (D-2) \, (\bar n \cdot l)^2\bigg \} \,,
\label{diagram a: hard region expression}
\end{eqnarray}
where the superscript ``$h$" denotes the hard contribution and we adopt the conventions
\begin{eqnarray}
l_{\perp}^2 \equiv g_{\perp}^{\mu \nu} \, l_{\mu}  \, l_{\nu}\,,  \qquad
g_{\perp}^{\mu \nu} \equiv g^{\mu \nu}-\frac{n^{\mu} \bar n^{\nu}}{2} -\frac{n^{\nu} \bar n^{\mu}}{2} \,.
\end{eqnarray}
The loop integrals can be evaluated directly, and we obtain
\begin{eqnarray}
\Pi_{\mu,  \, weak}^{(1), \, h}&=& \frac{\alpha_s \, C_F}{4 \, \pi} \, \tilde f_B(\mu) \, m_B \, 
\frac{\phi_{b \, \bar q}^{-}(\omega)}{\bar n \cdot p -\omega} \, 
\bigg \{ \bar n_{\mu}  \bigg [ {1 \over \epsilon^2} +
{1 \over \epsilon} \, \left ( 2 \, \ln {\mu \over  n \cdot p} + 1  \right ) + 2 \, \ln^2 {\mu \over  n \cdot p} \nonumber \\
&& + 2 \,\ln {\mu \over  m_b} -\ln^2 r - 2 \, {\rm Li_2} \left (- {\bar r \over r} \right )  
+{2-r \over r-1} \, \ln r +{\pi^2 \over 12} + 3 \bigg ] \nonumber \\
&& + n_{\mu}  \, \left [ {1 \over r-1} \, \left ( 1 +  {r \over \bar r}  \, \ln r  \right ) \right ]  \,  \bigg \} \,,
\label{diagram a: result of the hard region expression}
\end{eqnarray}
with $r=n \cdot p/m_b$ and $\bar r = 1-r$.

Along the same vein,  one can identify the hard-collinear contribution of $\Pi_{\mu,  \, weak}^{(1)}$ at leading power
\begin{eqnarray}
\Pi_{\mu,  \, weak}^{(1), \, hc}&=& i \, g_s^2 \, C_F \, \tilde f_B(\mu) \, m_B \, 
\frac{\phi_{b \bar q}^{-}(\omega)}{\bar n \cdot p -\omega} \int \frac{d^D \, l}{(2 \pi)^D} \, \nonumber \\
&&   \frac{2 \, m_b \, n \cdot (p+l)}{[ n \cdot (p+l) \,
\bar n \cdot (p-k+l) + l_{\perp}^2  + i 0][ m_b \, n \cdot l+ i 0] [l^2+i0]} \,,
\label{diagram a: hard-collinear region expression}
\end{eqnarray}
where the superscript ``$hc$" indicates the hard-collinear  contribution and the propagators have been expanded
systematically in the hard-collinear region.  Evaluating the integrals with the relations  yields
\begin{eqnarray}
\Pi_{\mu,  \, weak}^{(1), \, hc}&=& \frac{\alpha_s \, C_F}{4 \, \pi} \, \tilde f_B(\mu) \, m_B \, 
\frac{\phi_{b \bar q}^{-}(\omega)}{\omega - \bar n \cdot p} \,\, \bar n_{\mu}  \, \bigg [ {2 \over \epsilon^2} +
{2 \over \epsilon} \, \left (\ln {\mu^2 \over  n \cdot p \, (\omega - \bar n \cdot p)} + 1  \right )  \, \nonumber \\
&&  + \ln^2 {\mu^2 \over  n \cdot p \, (\omega - \bar n \cdot p)}
+ 2 \, \ln {\mu^2 \over  n \cdot p \, (\omega - \bar n \cdot p)}  
-{\pi^2 \over 6} + 4 \bigg ] \,.
\label{diagram a: result of the hard-collinear region expression}
\end{eqnarray}

Applying the method of regions we extract the soft contribution of $\Pi_{\mu, weak}^{(1)}$
\begin{eqnarray}
\Pi_{\mu,  \, weak}^{(1), \, s}&=& \frac{g_s^2 \, C_F}{2 \,( \bar n \cdot p -\omega) } \,
\int \frac{d^D \, l}{(2 \pi)^D} \, \frac{1}{[\bar n \cdot (p-k+l) + i 0][v \cdot l + i 0] [l^2+i0]}  \nonumber  \\
&& \bar d(k) \,\, \! \not n \,\,   \gamma_5  \,\,  \! \not {\bar n} \,\, \gamma_{\mu}  \,\,  b(p_b)  \,  \nonumber \\
&=& \frac{\alpha_s \, C_F}{4 \, \pi} \, \tilde f_B(\mu) \, m_B \, 
\frac{\phi_{b \bar d}^{-}(\omega)}{\bar n \cdot p -  \omega} \,\, \bar n_{\mu}  \,  \nonumber \\
&& \times \bigg [ {1 \over \epsilon^2} + {2 \over \epsilon} \, \ln {\mu \over  \omega - \bar n \cdot p}
+ 2 \, \ln^2 {\mu \over  \omega - \bar n \cdot p}    + {3 \, \pi^2 \over 4} \bigg ] \,,
\label{diagram a: result of the soft region expression}
\end{eqnarray}
where the superscript ``$s$" represents the soft contribution.

Now,  we compute the corresponding infrared subtraction term $\Phi_{b \bar d,  \, a}^{(1)} \otimes  T^{(0)}$
as displayed in figure \ref{fig: soft subtraction}(a).  With the Wilson-line Feynman rules, we obtain
\begin{eqnarray}
\Phi_{b \bar q, \, a}^{\alpha \beta\,, (1)} (\omega, \omega^{\prime})
&=& i \, g_s^2 \, C_F\, \int \frac{d^D \, l}{(2 \pi)^D} \,
\frac{1}{[\bar n \cdot l + i 0][v \cdot l + i 0] [l^2+i0]}  \nonumber  \\
&& \times [\delta(\omega^{\prime}-\omega-\bar n \cdot l)-\delta(\omega^{\prime}-\omega)] \,
[\bar d(k)]_{\alpha}  \, [b(v)]_{\beta} \,\,,
\label{effective diagram a: wave function}
\end{eqnarray}
from which we can derive the soft subtraction term
\begin{eqnarray}
\Phi_{b \bar q ,  \, a}^{(1)} \otimes T^{(0)}&=& \frac{g_s^2 \, C_F}{2 \,( \bar n \cdot p -\omega) } \,
\int \frac{d^D \, l}{(2 \pi)^D} \, \frac{1}{[\bar n \cdot (p-k+l) + i 0][v \cdot l + i 0] [l^2+i0]}  \nonumber  \\
&& \bar d(k) \,\, \! \not n \,\,   \gamma_5  \,\,  \! \not {\bar n} \,\, \gamma_{\mu}  \,\,  b(v)  \, \,,
\label{diagram a: soft subtraction}
\end{eqnarray}
We then conclude that
\begin{eqnarray}
\Pi_{\mu,  \, weak}^{(1), \, s}= \Phi_{b \bar q ,  \, a}^{(1)}  \otimes T^{(0)}\,
\end{eqnarray}
at leading power in $\Lambda / m_b$, which is an essential point to prove factorization
of the correlation function $\Pi_{\mu}$.

\begin{figure}[t]
\begin{center}
\includegraphics[width=0.8  \columnwidth]{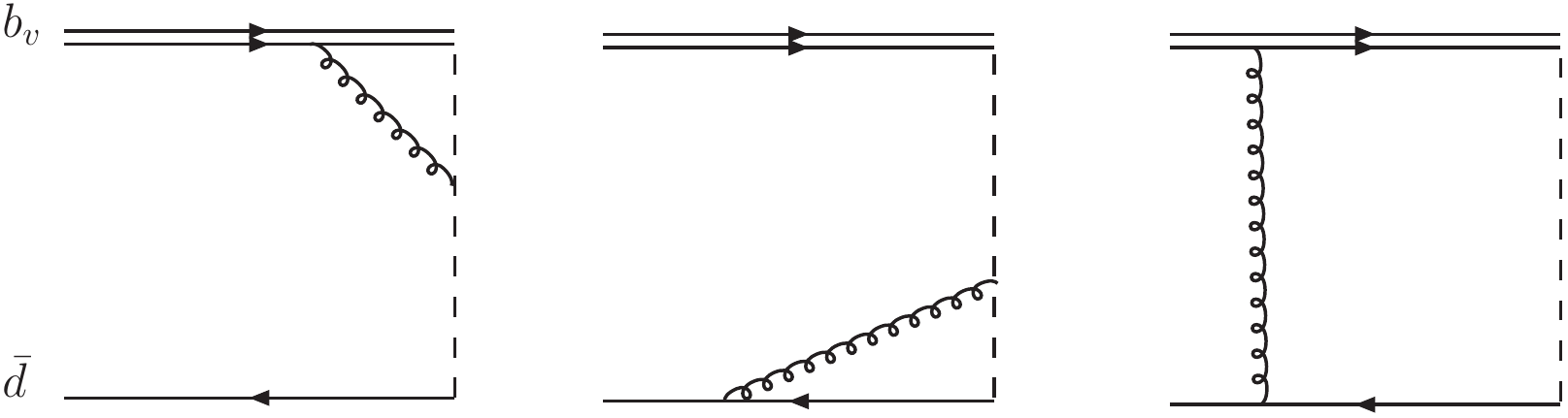}\\
\hspace{0.0 cm}(a) \hspace{3.5 cm} (b)\hspace{3.5 cm} (c)  \\
\vspace*{0.1cm}
\caption{One-loop diagrams for the $B$-meson DA $\Phi_{b \bar u}^{\alpha \beta}(\omega^{\prime})$. }
\label{fig: soft subtraction}
\end{center}
\end{figure}

\item{\bf Interpolation current vertex diagram}\\

Now we turn to compute the QCD correction to the Interpolation current vertex
(figure \ref{fig: NLO diagrams of the correlator}(b)).
The soft region contribution will generate a scaleless integral which vanishes in dimensional regularization
and the hard region induces a power-suppression factor $\lambda$ from the the spinor structure. Therefore, the contribution from the hard-collinear region will be the same as  the full QCD result. 
It is then straightforward to obtain contribution of current vertex 
\begin{eqnarray}
\Pi_{\mu,  \, P}^{(1)}&=& \Pi_{\mu,  \, P}^{(1),  \,hc}
= \frac{\alpha_s \, C_F}{4 \, \pi} \, \tilde f_B(\mu) \, m_B \, 
\frac{1}{\bar n \cdot p - \omega}  \,\,
\bigg \{ n_{\mu} \, \phi_{b \bar q}^{+}(\omega) \,
\left [ {\bar n \cdot p -\omega \over \omega} \, \ln {\bar n \cdot p -\omega \over \bar n \cdot p}  \right ] \nonumber \\
&& + \bar n_{\mu}  \,\,  \phi_{b \bar q}^{-}(\omega) \, \bigg [
\left ( {1 \over \epsilon} + \ln \left (- {\mu^2 \over p^2} \right )\right )\,
\left ( {2 \, \bar n \cdot p \over \omega} \, \ln {\bar n \cdot p -\omega \over \bar n \cdot p} + 1  \right )  \, \nonumber \\
&&  - {\bar n \cdot p \over \omega} \,  \ln {\bar n \cdot p - \omega \over  \bar n \cdot p}
\, \left ( \ln {\bar n \cdot p - \omega \over  \bar n \cdot p} + {2 \omega \over  \bar n \cdot p}  -4 \right ) + 4 \bigg ]  \bigg \}\,.
\label{diagram b: result of the hard-collinear region expression}
\end{eqnarray}
The soft contribution of $\Pi_{\mu,  \, P}^{(1)}$ vanishes in dimensional regularization, it can be proved  that
the precise cancelation  of $\Pi_{\mu,  \, P}^{(1),  \, s}$  and $\Phi_{b \bar q ,  \, b} \otimes T^{(0)}$ from figure \ref{fig: soft subtraction}(b) is independent of
regularization schemes.

\item{\bf Wave function renormalization}

The self-energy correction to the intermediate quark propagator
(figure \ref{fig: NLO diagrams of the correlator}(c)) is free of soft and collinear divergences and a straightforward
calculation gives
\begin{eqnarray}
\Pi_{\mu,  \, wfc}^{(1)}&=&  \frac{\alpha_s \, C_F}{4 \, \pi} \, \tilde f_B(\mu) \, m_B \, 
\frac{\phi_{b \bar q}^{-}(\omega)}{\bar n \cdot p - \omega} \,\, \bar n_{\mu}  \,\,
\bigg [{1 \over \epsilon} + \ln {\mu^2 \over n \cdot p \, (\omega-\bar n \cdot p)}+ 1 \bigg ] \,.
\label{diagram c: result}
\end{eqnarray}

For the external quark fields, the wave function renormalization of a massless quark
does not contribute to the matching coefficient when  dimensional regularization is applied to regularize
both ultraviolet and infrared divergences, i.e.,
\begin{eqnarray}
\Pi_{\mu,  \, dwf}^{(1)} -\Phi_{b \bar q,  \, dwf}^{(1)}  \otimes T^{(0)} =0 \,.
\label{u-quark wf in QCD}
\end{eqnarray}
The wave function renormalization of the $b$-quark is actually the match coefficient between the full QCD and HQET,
\begin{eqnarray}
\Pi_{\mu,  \, bwf}^{(1)} -\Phi_{b \bar q,  \, bwf}^{(1)}  \otimes T^{(0)} &=& - \frac{\alpha_s \, C_F}{8 \, \pi} \,
\bigg [{3 \over \epsilon} + 3 \, \ln {\mu^2 \over m_b^2} + 4 \bigg ] \, \Pi_{\mu}^{(0)} \,.
\label{matching coefficient from b-quark wf}
\end{eqnarray}

\item {\bf Box diagram}

The one-loop contribution to $\Pi_{\mu}$ from the box diagram (figure \ref{fig: NLO diagrams of the correlator}(d)) receives leading-power contribution from both the hard-collinear and the soft regions.
Evaluating the hard-collinear contribution of $\Pi_{\mu,  \, box}^{(1)}$ yields
\begin{eqnarray}
\Pi_{\mu,  \, box}^{(1), \, hc}&=& \frac{\alpha_s \, C_F}{4 \, \pi} \, \tilde f_B(\mu) \,  
\frac{m_B}{\omega} \, \bar n_{\mu}  \, 
\bigg \{\phi_{b \, \bar q}^{+}(\omega) \,  \bigg [ r\, \ln (1+\eta) \bigg ]
- 2 \, \phi_{b \, \bar q}^{-}(\omega) \,  \ln(1+\eta) \, \nonumber \\
&& \times \left [ {1 \over \epsilon}  + \ln {\mu^2 \over n \cdot p \, (\omega-\bar n \cdot p)}
+ {1 \over 2} \, \ln (1+\eta) + 1 \right ]  \,  \bigg \} \,,
\label{diagram d: result of the hard-collinear region}
\end{eqnarray}
with $\eta=-\omega/ \bar n \cdot p$.

Extracting the soft contribution of $\Pi_{\mu,  \, box}^{(1)}$ with the method of regions gives
\begin{eqnarray}
\Pi_{\mu,  \, box}^{(1), \,s}
&=& - \frac{g_s^2 \, C_F}{2}\,
\int \frac{d^D \, l}{(2 \pi)^D} \,   \frac{1}{[ v \cdot l+ i 0]
[\bar n \cdot (p-k+l) + i 0] [(k-l)^2+i0][l^2+i0]}  \nonumber  \\
&& \bar d(k)  \, \! \not v  \,  (\! \not k - \! \not l)  \,\,
\! \not n \,\gamma_5  \,\,  \! \not \bar n \,\, \gamma_{\mu} \, b(v)  \,.
\label{diagram d: soft region expression}
\end{eqnarray}
Now we compute the corresponding NLO contribution to the partonic LCDA
(figure \ref{fig: soft subtraction}(c))
\begin{eqnarray}
\Phi_{b \bar q,  \, c}^{\alpha \beta\,, (1)} (\omega, \omega^{\prime})
&=& - i \, g_s^2 \, C_F\, \int \frac{d^D \, l}{(2 \pi)^D} \,
\frac{1}{[(l-k)^2 + i 0][v \cdot l + i 0] [l^2+i0]}  \nonumber  \\
&& \times \delta(\omega^{\prime}-\omega+\bar n \cdot l) \,
[\bar d(k) \, \! \not v \,  (\! \not l - \! \not k) ]_{\alpha}  \, [b(v)]_{\beta} \,\,,
\label{effective diagram c: wave function}
\end{eqnarray}
from which one can deduce the soft subtraction term
\begin{eqnarray}
\Phi_{b \bar q,  \, c}^{(1)} \otimes T^{(0)} &=&  \frac{g_s^2 \, C_F}{2}\,
\int \frac{d^D \, l}{(2 \pi)^D} \,   \frac{1}{[ v \cdot l+ i 0]
[\bar n \cdot (p-k+l) + i 0] [(l-k)^2+i0][l^2+i0]}  \nonumber  \\
&& \bar d(k)  \, \! \not v  \,  (\! \not l - \! \not k)  \,\,
\! \not n \,\gamma_5  \,\,  \! \not \bar n \,\, \gamma_{\mu} \, b(v)  \,,
\label{diagram d: soft subtraction term}
\end{eqnarray}
which cancels out  the soft contribution of the correlation function $\Pi_{\mu,  \, box}^{(1) ,\, s}$ completely.
The absence of such soft contribution to the perturbative matching coefficient is particularly important for the box diagram,
since the relevant  loop integrals in the soft region depend  on {\it two} components of the soft spectator momentum
$\bar n \cdot k$ and $ v \cdot k$, and the light-cone OPE fails in the soft region.
\end{itemize}

After the results of all the diagrams have been obtained, the one-loop hard-scattering kernel of the correlation function $\Pi_{\mu}$ can be readily computed
from the matching condition in (\ref{matching condition of T1}) by collecting different pieces together
\begin{eqnarray}
\Phi_{b \bar q}^{(0)} \otimes  T^{(1)}  &=& \left [ \Pi_{\mu , \, weak}^{(1)}  + \Pi_{\mu , \, pion}^{(1)}
+ \Pi_{\mu , \, wfc}^{(1)}   + \Pi_{\mu , \, box}^{(1)}
+ \Pi_{\mu , \, bwf}^{(1)} + \Pi_{\mu , \, dwf}^{(1)} \right ] \nonumber \\
&& - \left [  \Phi_{b \bar q  , \, a}^{(1)} + \Phi_{b \bar q , \, b}^{(1)} + \Phi_{b \bar q , \, c}^{(1)}
+ \Phi_{b \bar q  , \, bwf}^{(1)}  + \Phi_{b \bar q  , \, dwf}^{(1)}  \right ]  \otimes  T^{(0)}  \, \nonumber \\
&=& \left [  \Pi_{\mu , \, weak}^{(1) , \, h}
+ \left ( \Pi_{\mu, \, bwf}^{(1)} -  \Phi_{b \bar q  , \, bwf}^{(1)} \right )\right ]
\nonumber \\
&& + \left [\Pi_{\mu , \, weak}^{(1) , \, hc} + \Pi_{\mu , \, P}^{(1) , \, hc} + \Pi_{\mu , \, wfc}^{(1) , \, hc}
+  \Pi_{\mu , \, box}^{(1) , \, hc}  \right ] \,, 
\label{schematic form of NLO hard kernel}
\end{eqnarray}
where the one-loop level hard function and the jet function can be extracted from the first and second square brackets of the second equality
respectively. Finally,   the  factorization formulae of $\Pi$ and $\widetilde{\Pi}$ are given by
\begin{eqnarray}
\Pi &=& \tilde{f}_B(\mu) \, m_B \sum \limits_{k=\pm} \,
C^{(k)}(n \cdot p, \mu) \, \int_0^{\infty} {d \omega \over \omega- \bar n \cdot p}~
J^{(k)}\left({\mu^2 \over n \cdot p \, \omega},{\omega \over \bar n \cdot p}\right) \,
\phi_{k}(\omega,\mu)  \,, \nonumber \\
\widetilde{\Pi} &=& \tilde{f}_B(\mu) \, m_B \sum \limits_{k=\pm} \,
\widetilde{C}^{(k)}(n \cdot p, \mu) \, \int_0^{\infty} {d \omega \over \omega- \bar n \cdot p}~
\widetilde{J}^{(k)}\left({\mu^2 \over n \cdot p \, \omega},{\omega \over \bar n \cdot p}\right) \,
\phi_{k}(\omega,\mu)  \,, \nonumber \\
\label{NLO factorization formula of correlator}
\end{eqnarray}
at leading power in $\Lambda/m_b$, where we keep  the factorization-scale dependence explicitly,
the hard coefficient functions are given by
\begin{eqnarray}
C^{(+)}  &=& \tilde{C}^{(+)}=1, \nonumber \\
C^{(-)} &=& \frac{\alpha_s \, C_F}{4 \, \pi}\, {1 \over \bar r} \,
\left [ {r \over \bar r} \, \ln r + 1 \right ]\,, \nonumber \\
\tilde{C}^{(-)} &=& 1 - \frac{\alpha_s \, C_F}{4 \, \pi}\,  \bigg [ 2 \, \ln^2 {\mu \over n \cdot p}
+ 5 \, \ln {\mu \over m_b} - \ln^2 r  - 2 \, {\rm Li_2} \left ( - {\bar r \over r} \right ) \nonumber \\
&& + {2-r \over r-1} \, \ln r  +  {\pi^2 \over 12} + 5 \bigg ] \,,
\label{results of hard coefficients}
\end{eqnarray}
and the jet functions are
\begin{eqnarray}
J^{(+)}&=& {1 \over r} \, \tilde{J}^{(+)}
= \frac{\alpha_s \, C_F}{4 \, \pi} \, \left (1- { \bar n \cdot p \over \omega } \right ) \,
\ln \left (1- { \omega  \over \bar n \cdot p } \right ) \,, \nonumber \\
\qquad  J^{(-)} &=& 1 \,, \nonumber \\
\tilde{J}^{(-)}&=& 1 + \frac{\alpha_s \, C_F}{4 \, \pi} \,
\bigg [ \ln^2 { \mu^2 \over  n \cdot p (\omega- \bar n \cdot p) }
- 2 \ln {\bar n \cdot p -\omega \over \bar n \cdot p } \, \ln { \mu^2 \over  n \cdot p (\omega- \bar n \cdot p) }
\,  \nonumber \\
&& - \ln^2 {\bar n \cdot p -\omega \over \bar n \cdot p }
- \left ( 1 +  {2 \bar n \cdot p \over \omega} \right )  \ln {\bar n \cdot p -\omega \over \bar n \cdot p }
-{\pi^2 \over 6} -1 \bigg ] \,.
\label{results of jet functions}
\end{eqnarray}

From one-loop calculation, one can obtained the anomalous dimension of the hard function, jet function and the soft contribution, then the factorization-scale independence of $\Pi$ and $\widetilde{\Pi}$ can demonstrate explicitly. It is straightforward to write down the following evolution
equations
\begin{eqnarray}
&& {d \over d \ln \mu} \widetilde{C}^{(-)}(n \cdot p, \mu) = - \frac{\alpha_s \, C_F}{4 \, \pi}
\left [ \Gamma_{\rm cusp}^{(0)} \ln { \mu \over n \cdot p} + 5\right ]  \widetilde{C}^{(-)}(n \cdot p, \mu)\,,
\label{RGE of tildeC1}\\
&& {d \over d \ln \mu} \tilde{J}^{(-)}\left({\mu^2 \over n \cdot p \, \omega},{\omega \over \bar n \cdot p}\right)
= \frac{\alpha_s \, C_F}{4 \, \pi} \left [ \Gamma_{\rm cusp}^{(0)} \ln { \mu^2 \over n \cdot p\, \omega} \right ]
\tilde{J}^{(-)}\left({\mu^2 \over n \cdot p \, \omega},{\omega \over \bar n \cdot p}\right) \nonumber \\
&& \hspace{3 cm} +  \frac{\alpha_s \, C_F}{4 \, \pi} \, \int_0^{\infty} \, d \omega^{\prime}  \, \omega \,\,
\Gamma(\omega,\omega^{\prime},\mu) \,\,
\tilde{J}^{(-)} \left({\mu^2 \over n \cdot p \, \omega^{\prime}},{\omega^{\prime} \over \bar n \cdot p}\right)  \,,  \\
&& {d \over d \ln \mu} \left [ \tilde{f}_B(\mu) \, \phi_-(\omega, \mu) \right ] =  - \frac{\alpha_s \, C_F}{4 \, \pi}
\left [ \Gamma_{\rm cusp}^{(0)} \ln { \mu \over \omega} - 5\right ] \left [ \tilde{f}_B(\mu) \, \phi_-(\omega, \mu) \right ]  \nonumber \\
&& \hspace{3.5 cm} -  \frac{\alpha_s \, C_F}{4 \, \pi} \, \int_0^{\infty} \, d \omega^{\prime}  \, \omega \,\,
\Gamma(\omega,\omega^{\prime},\mu) \,\, \left [ \tilde{f}_B(\mu) \, \phi_-(\omega^{\prime}, \mu) \right ] \,,
\end{eqnarray}
where the function last equation is actually the Lange-Nuebert equation(except for the anomalous dimension of the decay constant $f_B$) for $\phi_-$, and $\Gamma$ has been given in the previous section.
The renormalization kernel of $\phi_-(\omega, \mu)$ at one-loop level was first computed in
\cite{Bell:2008er} and then confirmed in \cite{DescotesGenon:2009hk}.
With the evolution equations displayed above, it is evident that
\begin{eqnarray}
{d \over d \ln \mu} \left [\Pi(n \cdot p, \bar n \cdot p) \,,
\widetilde{\Pi}(n \cdot p, \bar n \cdot p)  \right ]
= {\cal O}(\alpha_s^2).
\end{eqnarray}

One cannot avoid the  large logarithms of order $\ln (m_b/\Lambda_{\rm QCD})$ in the hard functions, the jet functions, $\tilde{f}_B(\mu)$ and the $B$-meson DAs concurrently by choosing a common value of $\mu$, and the large logarithms have to be resummed by solving the three RG equations shown above. The solution to evolution equation of the LCDA has been discussed detailed in the previous section, in addtion,
the hadronic scale entering the initial conditions of the $B$-meson LCDAs  $\mu_0\approx 1.0 \,\, {\rm GeV}$ is quite close
to the hard-collinear scale $\mu_{hc} \simeq  \sqrt{m_b \, \Lambda_{\rm QCD}} \approx 1.5 \,\, {\rm GeV}$, it is not important phenomenologically to sum
logarithms of $\mu_{hc}/\mu_0$ \cite{Beneke:2011nf}.
To achieve NLL resummation of large logarithms in the hard coefficient $\widetilde{C}^{(-)}$
we need to generalize  the RG equation (\ref{RGE of tildeC1}) to
\begin{eqnarray}
{d \over d \ln \mu} \widetilde{C}^{(-)}(n \cdot p, \mu) &=&
\left [ - \Gamma_{\rm cusp}(\alpha_s) \ln { \mu \over n \cdot p} + \gamma(\alpha_s) \right ]  \widetilde{C}^{(-)}(n \cdot p, \mu)\,,
\nonumber \\
{d \over d \ln \mu} \, \tilde{f}_B(\mu) &=&\tilde{\gamma}(\alpha_s)\, \tilde{f}_B(\mu) \,.
\label{general RGE of tildeC1}
\end{eqnarray}
The  cusp anomalous dimension at the three-loop order and the remanning anomalous dimension
$\gamma(\alpha_s)$  determining  renormalization of the SCET heavy-to-light current at two loops
will enter $U_1(n \cdot p,\mu_{h1},\mu )$ at NLL accuracy. The manifest expressions of $\Gamma_{\rm cusp}^{(i)}$,
$\gamma^{(i)}$ and $\beta_i$ can be found in \cite{Beneke:2011nf} and references therein
\footnote{Note that there is a factor $C_F$ difference of  our conventions of $\Gamma_{\rm cusp}^{(i)}$ and
$\gamma^{(i)}$ compared with   \cite{Beneke:2011nf}.},
the evolution function $U_1(n \cdot p,\mu_{h1},\mu )$ can be read from Eq. (A.3) in \cite{Beneke:2011nf}
with the replacement rules $E_{\gamma} \to {n \cdot p / 2}$ and $\mu_h \to \mu_{h1}$.
Because the hard scale $\mu_{h1} \sim  n \cdot p$ in the hard function $\widetilde{C}^{(-)}(n \cdot p, \mu)$ differs from
the one $\mu_{h2} \sim  m_b$ in $\tilde f_B(\mu)$,  the resulting evolution functions due to running of the renormalization scale
from $\mu_{h1}$($\mu_{h2}$) to $\mu_{hc}$ in $\widetilde{C}^{(-)}(n \cdot p, \mu)$ ($\tilde f_B(\mu)$) are
\begin{eqnarray}
\widetilde{C}^{(-)}(n \cdot p, \mu) &=& U_1(n \cdot p,\mu_{h1},\mu ) \, \widetilde{C}^{(-)}(n \cdot p, \mu_{h1}) \,, \nonumber \\
\tilde f_B(\mu) &=& U_2(\mu_{h2},\mu ) \, \tilde f_B(\mu_{h2}) \,.
\end{eqnarray}

The final factorization formulae of $\Pi$ and $\widetilde{\Pi}$ with RG improvement at NLL accuracy can be written as
\begin{eqnarray}
\Pi &=& m_B    \, \left [U_2(\mu_{h2},\mu ) \, \tilde{f}_B(\mu_{h2}) \right ]
\int_0^{\infty} {d \omega \over \omega- \bar n \cdot p}~
J^{(+)}\left({\mu^2 \over n \cdot p \, \omega},{\omega \over \bar n \cdot p}\right) \,
\phi_{+}(\omega,\mu)  \, \nonumber \\
&& + m_B  \,\left [U_2(\mu_{h2},\mu ) \, \tilde{f}_B(\mu_{h2}) \right ] \, C^{(-)}(n \cdot p, \mu) \,
\int_0^{\infty} {d \omega \over \omega- \bar n \cdot p}~ \,
\phi_{-}(\omega,\mu)  \,, \nonumber \\
\widetilde{\Pi} &=& m_B  \, \left [U_2(\mu_{h2},\mu ) \, \tilde{f}_B(\mu_{h2}) \right ]
\int_0^{\infty} {d \omega \over \omega- \bar n \cdot p}~
\widetilde{J}^{(+)}\left({\mu^2 \over n \cdot p \, \omega},{\omega \over \bar n \cdot p}\right) \,
\phi_{+}(\omega,\mu)  \, \nonumber \\
&& + m_B   \, \left [U_1(n \cdot p,\mu_{h1},\mu ) \, U_2(\mu_{h2},\mu ) \right ] \,
\left [ \tilde{f}_B(\mu_{h2}) \, \widetilde{C}^{(-)}(n \cdot p, \mu_{h1})  \right ] \,  \nonumber \\
&& \hspace{0.3 cm} \times  \int_0^{\infty} {d \omega \over \omega- \bar n \cdot p}~
\widetilde{J}^{(-)}\left({\mu^2 \over n \cdot p \, \omega},{\omega \over \bar n \cdot p}\right) \,
\phi_{-}(\omega,\mu)  \,,
\label{resummation improved factorization formula}
\end{eqnarray}
where $\mu$ should be taken as a hard-collinear scale of order $\sqrt{m_b \, \Lambda}$.

Since the one-loop parton level result of the correlation function has been obtained, we are ready to construct the sum rules of $f_{B\to P}^{+}(q^2)$ and $f_{B\to P}^{0}(q^2)$ including
the  radiative corrections at ${\cal O}(\alpha_s)$. In order to achieve this target, one must expression the one-loop correlation function in term of the dispersion integral. 
Using quark-hadron duality assumption and the Borel transform, we obtain the form factors for the $B\to\pi$ transition
\begin{eqnarray}
&& f_{\pi} \,\, e^{-m_{\pi}^2/(n \cdot p \, \omega^{\pi}_M)} \,\,\,
\left \{ \frac{n \cdot p} {m_B} \, f_{B\to \pi}^{+}(q^2)
\,, \,\,\,   f_{B\to \pi}^{0}(q^2)  \right \}  \,  \nonumber \\
&& = \left [U_2(\mu_{h2},\mu ) \, \tilde{f}_B(\mu_{h2}) \right ]
\,\,\, \int_0^{\omega^{\pi}_s} \,\, d \omega^{\prime} \,\, 
\, e^{-\omega^{\prime} / \omega^{\pi}_M}  \, \bigg [  r\, \phi_{B, \rm eff}^{+}(\omega^{\prime}, \mu) \nonumber \\
&& \hspace{0.4 cm} + \left [ U_1(n \cdot p,\mu_{h1},\mu ) \, \widetilde{C}^{(-)}(n \cdot p, \mu_{h1})  \right ]
\, \phi_{B, \rm eff}^{-}(\omega^{\prime}, \mu)  \nonumber \\
&& \hspace{0.4 cm}  \pm \,\,\, \frac{n \cdot p -m_B} {m_B}   \,\,\,
\left ( \phi_{B, \rm eff}^{+}(\omega^{\prime}, \mu)
+ C^{(-)}(n \cdot p, \mu) \, \phi_{B}^{-}(\omega^{\prime}, \mu) \right )  \bigg ] \,,
\label{NLO sum rules of form factors}
\end{eqnarray}
where the functions $\phi_{B, \rm eff}^{\pm}(\omega^{\prime}, \mu)$ are defined as
\begin{eqnarray}
\phi_{B, \rm eff}^{+}(\omega^{\prime}, \mu) &=&  \frac{\alpha_s \, C_F}{4 \, \pi} \,\,
\int_{\omega^{\prime}}^{\infty} \,\, {d \omega \over \omega} \,\, \phi_{+}(\omega, \mu) \,\,\,,
\label{effective B-meson plus}  \\
\phi_{B, \rm eff}^{-}(\omega^{\prime}, \mu) &=& \phi_{-}(\omega^{\prime}, \mu)
+  \frac{\alpha_s \, C_F}{4 \, \pi} \,\, \bigg \{ \int_0^{\omega^{\prime}} \,\, d \omega \,\,\,
\left [ {2 \over \omega - \omega^{\prime}}  \,\,\, \left (\ln {\mu^2 \over n \cdot p \, \omega^{\prime}}
- 2 \, \ln {\omega^{\prime} - \omega \over \omega^{\prime}} \right )\right ]_{\oplus}\phi_{-}(\omega, \mu) \,\, \nonumber \\
&& \hspace{-1.5cm}
- \int_{\omega^{\prime}}^{\infty} \! d \omega
\bigg [ \ln^2 {\mu^2 \over n \cdot p \, \omega^{\prime}} - \left ( 2 \, \ln {\mu^2 \over n \cdot p \, \omega^{\prime}}  + 3 \right )
\ln {\omega - \omega^{\prime} \over \omega^{\prime}}  + 2 \, \ln {\omega \over \omega^{\prime}}   + {\pi^2 \over 6} - 1 \bigg ]
\, {d \phi_{-}(\omega, \mu) \over d \omega}  \bigg \} \,.
\label{effective B-meson minus}
\end{eqnarray}
The other $B\to P$ form factors could be calculated in the same way.

\subsection{The $B \to V$ form factors with SCET improved LCSR}

In this subsection we will calculate the NLO corrections to the SCET form factors $\xi_i$ and $\Xi_i$.
The computation of the form factor $\xi_\|$ will be presented in detail, the calculation of the other form factors, which can be derived in a similar procedure, will be neglected.

We start with the the SCET vacuum-to-$B$-meson correlation function (\ref{correlation function for xi-L})
\begin{eqnarray}
\Pi_{\nu, \|}(p, q) = \int d^4 x \, e^{i p \cdot x} \, \langle 0 |
{\rm T}  \left \{j_{\nu}(x),  \,\, \left (\bar \xi \, W_c \right)(0) \,
\gamma_5 \, h_v(0) \,    \right \}   | \bar B_v \rangle \,.
\end{eqnarray}
It is straightforward to write down the leading-power contribution to the correlation function
\begin{eqnarray}
&& \Pi_{\nu, \|}(p, q) \nonumber \\
&& = \int d^4 x \, e^{i p \cdot x} \,
 \langle 0 | {\rm T}  \left \{ j_{\xi q_s, \parallel \, \nu}^{(2)}(x),  \,\, \left (\bar \xi \, W_c \right)(0) \,
\gamma_5 \, h_v(0) \,    \right \}   | \bar B_v \rangle \nonumber \\
&& + \int d^4 x \, e^{i p \cdot x} \, \int d^4 y \,
 \langle 0 | {\rm T}  \left \{ j_{\xi \xi, \nu}^{(0)}(x),  \,\, i \, {\cal L}_{\xi q_s}^{(2)}(y),  \,\,
\left (\bar \xi \, W_c \right)(0) \, \gamma_5 \, h_v(0) \,    \right \}   | \bar B_v \rangle  \nonumber \\
&& + \int d^4 x \, e^{i p \cdot x} \, \int d^4 y \, \int d^4 z \,
 \langle 0 | {\rm T}  \left \{ j_{\xi \xi, \nu}^{(0)}(x),  \,\, i \, {\cal L}_{\xi q_s}^{(1)}(y),  \,\,
i \, {\cal L}_{\xi m}^{(1)}(z),  \,\,
\left (\bar \xi \, W_c \right)(0) \, \gamma_5 \, h_v(0) \,    \right \}   | \bar B_v \rangle  \nonumber \\
&& \equiv \Pi_{\nu, \|}^{A}(p, q)  + \Pi_{\nu, \|}^{B}(p, q) + \Pi_{\nu, \|}^{C}(p, q) \,,
\label{def：xi-L-A-B-C}
\end{eqnarray}
where the third term $\Pi_{\nu, \|}^{C}$ takes into account the light-quark mass effect.
The multipole expanded SCET Lagrangian up to the ${\cal O}(\lambda^2)$ accuracy \cite{Beneke:2002ni}
have been derived with the position-space formalism \cite{Beneke:2002ph}
\begin{eqnarray}
{\cal L}_{\xi}^{(0)} &=& \bar \xi \, \left ( i \, \bar n \cdot D
+ i \not \! \! D_{\perp c} \,\, {1 \over i \, n \cdot D_c} \, i \not \! \! D_{\perp c} \right ) \,
{\not \! n \over 2 }  \,\, \xi  \,,  \nonumber \\
{\cal L}_{\xi m}^{(1)} &=& m \,\, \bar \xi \, \left [ i \not \! \! D_{\perp c},  \, {1 \over i \, n \cdot D_c}   \right ] \,
{\not \! n \over 2 }  \,\, \xi  \,, \nonumber \\
{\cal L}_{\xi m}^{(2)} &=& -m^2 \,\, \bar \xi \,  {1 \over i \, n \cdot D_c}  \,
{\not \! n \over 2 }  \,\, \xi  \nonumber   \,, \\
{\cal L}_{\xi q_s}^{(1)} &=& \bar q_s \,  W_c^{\dag} \,\,  i \not \! \! D_{\perp c}\,\, \xi
- \bar \xi \,\,   i \not \! \! \overleftarrow{D}_{\perp c} \,\,  W_c \, q_s,   \nonumber   \\
{\cal L}_{\xi q_s}^{(2)} &=& \bar q_s \,  W_c^{\dag} \,\,   \left ( i \, \bar n \cdot D
+ i \not \! \! D_{\perp c} \,\, {1 \over i \, n \cdot D_c} \, i \not \! \! D_{\perp c} \right ) \,\,
{\not \! n \over 2 }  \,\, \xi   \nonumber \\
&& -  \, \bar \xi \, {\not \! n \over 2 } \,\,  \left ( i \, \bar n \cdot \overleftarrow{D}
+ \, i \not \! \! \overleftarrow{D}_{\perp c} \,\, {1 \over i \, n \cdot \overleftarrow{D}_c} \,
i \not \! \! \overleftarrow{D}_{\perp c} \right ) \,\,
W_c \,\, q_s   \nonumber \\
&& + \,  \bar q_s  \, \overleftarrow{D}_s^{\mu} \, x_{\perp \mu} \,  W_c^{\dag} \,\,  i \not \! \! D_{\perp c}\,\, \xi
- \bar \xi  \,  i \not \! \! \! \overleftarrow{ D}_{\perp c} \,\,  W_c \, x_{\perp \mu}  \, D_s^{\mu} \,   q_s \,.
\end{eqnarray}
In the following we aim at calculating the jet function in the following factorization formula
$\Pi_{\nu, \|}^{i}$ as defined in (\ref{def：xi-L-A-B-C}) onto ${\rm SCET}_{\rm II}$
\begin{eqnarray}
\Pi_{\nu, \|}^{i}(p, q) = {\tilde{f}_B(\mu) \, m_B \over 2}\,
\sum_{m = \pm } \, \int_0^{+\infty} \, d \omega \, J_{\|, m}^{i}
\left ({\mu^2 \over n \cdot p \, \omega}, {\omega \over \bar n \cdot p} \right )  \,
\phi_{m}(\omega, \mu) \,\, \bar n_{\nu},  \,\, \, (i=A, B, C)
\label{SCET factorization formula L}
\end{eqnarray}
at one-loop accuracy.

Firstly we calculate the jet function $J_{\|, m}^{A} $ entering the SCET factorization formula (\ref{SCET factorization formula L}) by investigating the following matrix element
\begin{eqnarray}
F_{\|}^{A} = \int d^4 x \, e^{i p \cdot x} \,
\left \langle 0 \left | {\rm T}  \left \{ \bar q_s(x) \, Y_s \,  {\not \! n \over 2} \,  W_c^{\dagger} \, \xi(x),
\,\, \left (\bar \xi \, W_c \right)(0) \, \gamma_5 \, h_v(0) \,    \right \}   \right | \bar q_s(k) \, h_v \right \rangle \,.
\label{correlator: F-L-A}
\end{eqnarray}
At tree level we have
\begin{eqnarray}
F_{\|, \, \rm LO}^{A} =  - {i \over \bar n \cdot p - \omega^{\prime} + i0} \ast
\langle O_{\|, \,-} (\omega, \omega^{\prime})  \rangle^{(0)}  \,,
\end{eqnarray}
where the light-cone  matrix element $\langle O_{\|, \, -} (\omega, \omega^{\prime}) \rangle$
is defined as
\begin{eqnarray}
\langle O_{\|, \, -} (\omega, \omega^{\prime})  \rangle
= \langle 0 |O_{\|, \, -} (\omega^{\prime}) | \bar q_s(k) \, h_v  \rangle
=\bar q_s(k) \, {\not \! n \over 2} \, \gamma_5 \,  h_v  \,\,
\delta(\omega - \omega^{\prime}) + {\cal O}(\alpha_s),
\end{eqnarray}
and the HQET operator $O_{\|, \, \mp} (\omega^{\prime})$  in the  momentum space reads
\begin{eqnarray}
O_{\|, \, \mp} (\omega^{\prime}) = {1 \over 2 \, \pi} \, \int d t \, e^{i \, t \, \omega^{\prime}} \,
\left ( \bar q_s Y_s \right )(t \, \bar n) \, \,  \, \gamma_5 \,\,\left\{{\not \! n \over 2},{\not \! \bar n \over 2} \right\}
\left (Y_s^{\dag} h_v \right ) (0)\,.
\end{eqnarray}
Remembering the definition of the LCDA $\phi_-$, we can obtain the jet functions at tree level as follows
\begin{eqnarray}
J_{\|, \, -}^{A, \, (0)} = {1 \over \bar n \cdot p - \omega^{\prime} + i0}\,,
\qquad J_{\|, \, +}^{A, \, (0)} = 0 \,.
\end{eqnarray}

\begin{figure}
\begin{center}
\includegraphics[width=0.85 \columnwidth]{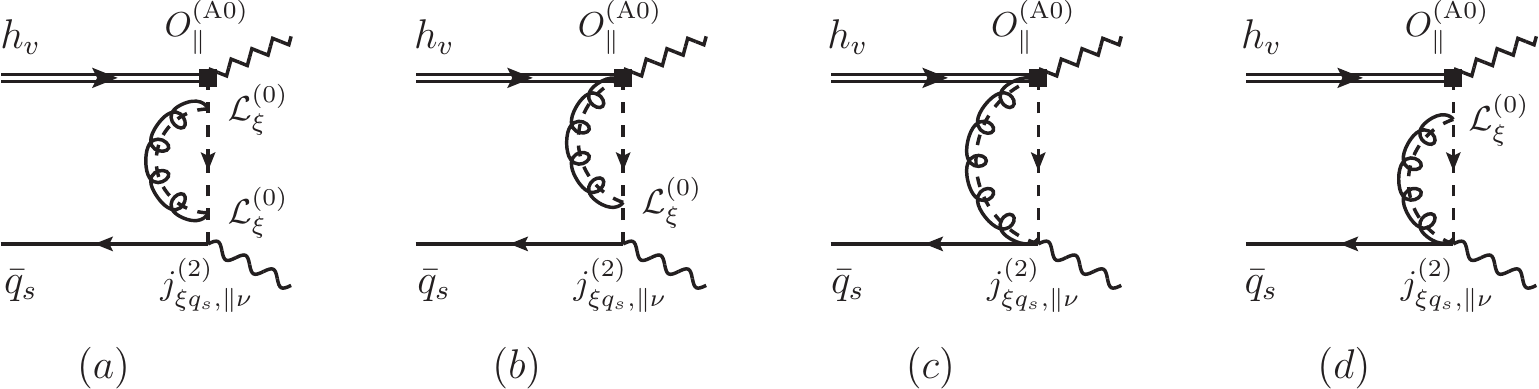}
\vspace*{0.1cm}
\caption{Diagrammatical representation of the vacuum-to-$B$-meson correlation function
$\Pi_{\nu, \|}^{A}(p, q)$  defined with  the ${\rm A0}$-type SCET operator
$O_{\|}^{\rm {(A0)}}= \left (\bar \xi \, W_c \right) \, \gamma_5 \, h_v$
and the power suppressed interpolating current $j_{\xi q_s, \parallel \, \nu}^{(2)}$ at  one loop.}
\label{fig: correlator for F-L-A}
\end{center}
\end{figure}
We proceed to determine the NLO contribution to the jet function $ J_{\|, \pm}^{A}$, the corresponding one-loop ${\rm SCET}_{\rm I}$ diagrams are
presented in figure \ref{fig: correlator for F-L-A} with the
subleading power SCET Feynman rules collected in \cite{Beneke:2018rbh}.
The self-energy correction to the hard-collinear quark propagator displayed
in figure \ref{fig: correlator for F-L-A}(a)
can be readily written as \cite{Wang:2015ndk}
\begin{eqnarray}
F_{\|, \, {\rm NLO}}^{A, (a)} = -{\alpha_s \, C_F \over 4 \, \pi} \,
\left [ {1 \over \epsilon}  + \ln {\mu^2 \over n \cdot p \, (\omega- \bar n \cdot p) - i 0} + 1 \right ]  \,
F_{\|, \, {\rm LO}}^{A} \,.
\end{eqnarray}
Diagram (c) of figure \ref{fig: correlator for F-L-A}
yields vanishing contribution due to $n^2=0$ where $n$ is from the Wilson line.
One can further verify that the hard-collinear corrections displayed in
the diagrams (b) and (d) of figure \ref{fig: correlator for F-L-A}
give rise to the identical  results
\begin{eqnarray}
F_{\|, \, {\rm NLO}}^{A, (b)} &=& F_{\|, \, {\rm NLO}}^{A, (d)}
= - {2 \, g_s^2 \, C_F \over \bar n \cdot p - \omega} \,\,
\bar q_s(k) \, {\not \! n \over 2} \, \gamma_5 \,  h_v  \, \nonumber \\
&& \times \, \int {d^D l \over  (2 \pi)^D} \,
{n \cdot (p+l) \over [n \cdot (p+l) \, \bar n \cdot (p-k+l) + l_{\perp}^2 + i 0]
[ n \cdot l + i 0] [l^2 + i 0] }  \,,
\end{eqnarray}
which can be evaluated straightforwardly with dimensional regularization scheme
\begin{eqnarray}
F_{\|, \, {\rm NLO}}^{A, (b)} &=& F_{\|, \, {\rm NLO}}^{A, (d)}
= {\alpha_s \, C_F \over 2 \, \pi} \,\,
\bigg  \{ {1 \over \epsilon^2}   + {1 \over \epsilon} \,
\left [  \ln {\mu^2 \over n \cdot p \, (\omega- \bar n \cdot p)} + 1 \right ]
+ {1 \over 2} \, \ln^2 {\mu^2 \over n \cdot p \, (\omega- \bar n \cdot p)} \nonumber \\
&& +  \, \ln {\mu^2 \over n \cdot p \, (\omega- \bar n \cdot p)}
- {\pi^2 \over 12} + 2  \bigg \}   \,\, F_{\|, \, {\rm LO}}^{A}  \,.
\end{eqnarray}
Adding up different pieces together leads to the jet functions at the one-loop accuracy
\begin{eqnarray}
J_{\|, \, -}^{A, \, (1)} &=&  J_{\|, \, -}^{A, \, (0)} \,
\bigg \{ 1 +  {\alpha_s \, C_F \over 4 \, \pi}  \,
\bigg  [ {4 \over \epsilon^2}   + {1 \over \epsilon} \,
\left ( 4 \,  \ln {\mu^2 \over n \cdot p \, (\omega- \bar n \cdot p) } + 3  \right )
+ 2 \, \ln^2 {\mu^2 \over n \cdot p \, (\omega- \bar n \cdot p)} \nonumber \\
&& + \, 3 \, \ln {\mu^2 \over n \cdot p \, (\omega- \bar n \cdot p)}
- {\pi^2 \over 3} + 7  \bigg ]  \bigg \} \,, \nonumber \\
 J_{\|, \, +}^{A, \, (1)} &=& 0 \,,
\label{jet functions: L-A}
\end{eqnarray}
which are in precise agreement with the results presented in \cite{DeFazio:2007hw}.

\begin{figure}[!t]
\begin{center}
\includegraphics[width=0.8 \columnwidth]{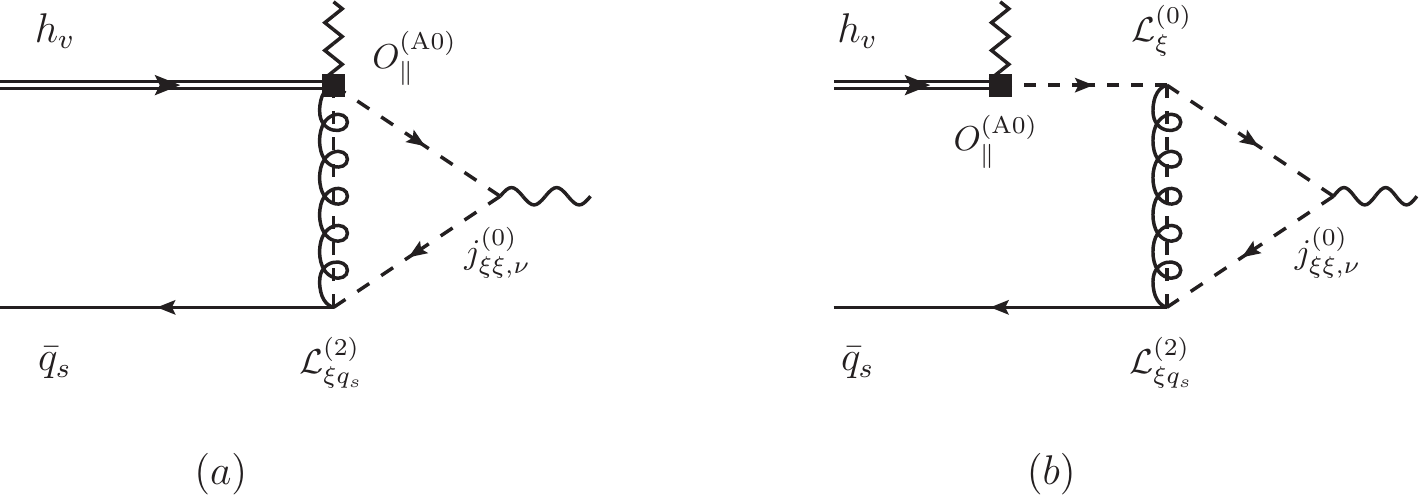}
\vspace*{0.1cm}
\caption{Diagrammatical representation of the vacuum-to-$B$-meson correlation function
$\Pi_{\nu, \|}^{B}(p, q)$  defined with  the ${\rm A0}$-type SCET operator
$O_{\|}^{\rm {(A0)}}= \left (\bar \xi \, W_c \right) \, \gamma_5 \, h_v$,
the leading power interpolating current $j_{\xi \xi, \nu}^{(0)}$
and the subleading power SCET Lagrangian ${\cal L}_{\xi q_s}^{(2)}$.}
\label{fig: correlator for F-L-B}
\end{center}
\end{figure}
\begin{figure}[!t]
\begin{center}
\includegraphics[width=0.4 \columnwidth]{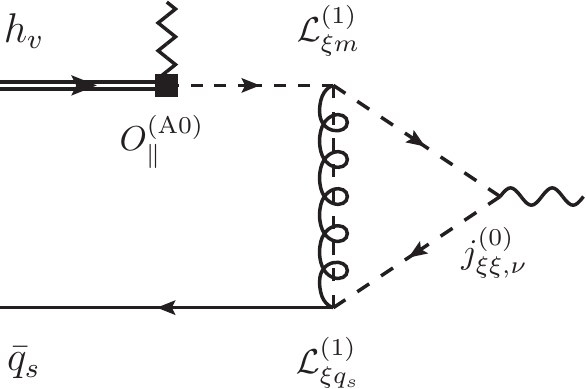}
\vspace*{0.1cm}
\caption{Diagrammatical representation of the vacuum-to-$B$-meson correlation function
$\Pi_{\nu, \|}^{B}(p, q)$  defined with  the ${\rm A0}$-type SCET operator
$O_{\|}^{\rm {(A0)}}= \left (\bar \xi \, W_c \right) \, \gamma_5 \, h_v$,
the leading power interpolating current $j_{\xi \xi, \nu}^{(0)}$
and the subleading power SCET Lagrangians ${\cal L}_{\xi q_s}^{(1)}$
and ${\cal L}_{\xi m}^{(1)}$.}
\label{fig: correlator for F-L-C}
\end{center}
\end{figure}

The jet function $J_{\|, \, \pm}^{B}$  and $J_{\|, \, \pm}^{C}$ appear only at loop level,  and they can be determined  with similar method with $J_{\|, \, \pm}^{A}$. $J_{\|, \, \pm}^{B}$ and $J_{\|, \, \pm}^{C}$ can be obtained through diagrams in figure \ref{fig: correlator for F-L-B} and the diagram in figure \ref{fig: correlator for F-L-C} with the SCET Feynman rules, respectively.
We collect the corresponding jet functions here
\begin{eqnarray}
J_{\|, \, -}^{B} &=&  {\alpha_s \, C_F \over 4 \, \pi}  \, J_{\|, \, -}^{A, \, (0)} \,
\bigg \{   - {2 \over \epsilon^2}   + {1 \over \epsilon} \,
\left [ - 2 \,  \ln \left ( {\mu^2 \over n \cdot p \, (\omega- \bar n \cdot p) } \right )  - 2 \, \ln (1+ \eta) - 3  \right ]
\nonumber \\
&& - \,  \ln^2 \left ( {\mu^2 \over n \cdot p \, (\omega- \bar n \cdot p) } \right )
+  \ln \left ( {\mu^2 \over n \cdot p \, (\omega- \bar n \cdot p) } \right ) \,
\left [ - 2 \, \ln (1+ \eta) - 3 \right ] - \ln^2 (1+\eta) \nonumber \\
&& + \,  \left ({2 \over \eta} -1 \right ) \, \ln (1+ \eta)
+ {\pi^2 \over 6} - 8  \bigg \} \,, \nonumber \\
 J_{\|, \, +}^{B} &=& 0 \,,
 \nonumber \\
 J_{\|, \, +}^{C} &=&  -  {m \over \omega}  \,
{1 \over \bar n \cdot p - \omega + i0} \,  {\alpha_s \, C_F \over 4 \, \pi}  \,
\ln \left ( {\bar n \cdot p - \omega \over \bar n \cdot p} \right ) \,, \nonumber \\
J_{\|, \, -}^{C} &=& 0 \,,
 \label{jet functions: L-C}
\end{eqnarray}

Plugging the obtained jet functions into the factorization formula
(\ref{SCET factorization formula L}) and employing the decomposition of  $\Pi_{\nu, \|}$
defined in (\ref{def：xi-L-A-B-C}) yields
\begin{eqnarray}
\Pi_{\nu, \|}(p, q) &=& {\tilde{f}_B(\mu) \, m_B \over 2}\,
\, \int_0^{+\infty} \, {d \omega \over \bar n \cdot p - \omega + i0} \,
\bigg \{  \left [ 1 + {\alpha_s \, C_F \over 4 \, \pi}  \, \hat{J}_{\|,-}^{(\rm A0)}
\left ({\mu^2 \over n \cdot p \, \omega}, {\omega \over \bar n \cdot p} \right ) \right ] \,
\phi_{-}(\omega, \mu) \,\,  \nonumber \\
&&  + \left [ \, {\alpha_s \, C_F \over 4 \, \pi}  \, \hat{J}_{\|,+}^{(m)}
\left ({\mu^2 \over n \cdot p \, \omega}, {\omega \over \bar n \cdot p} \right ) \right ] \,
\phi_{+}(\omega, \mu)  \bigg \} \,\,  \bar n_{\nu} \,,
\label{Final SCET factorization formula for Pi-L}
\end{eqnarray}
where the normalized one-loop jet functions $\hat{J}_{\|,-}^{(\rm A0)}$ and $\hat{J}_{\|,+}^{(m)}$ read
\begin{eqnarray}
\hat{J}_{\|,-}^{(\rm A0)} &=&  \ln^2 \left ( {\mu^2 \over n \cdot p \, (\omega- \bar n \cdot p) } \right )
- 2 \, \ln \left ( {\mu^2 \over n \cdot p \, (\omega- \bar n \cdot p) } \right ) \, \ln(1+\eta)
-\ln^2 (1+\eta) \nonumber \\
&& + \left ({2 \over \eta} -1 \right ) \,\ln(1+\eta)  - {\pi^2 \over 6}  - 1 \,, \nonumber \\
\hat{J}_{\|,+}^{(m)} &=&  - {m \over \omega}  \, \ln \left ( {\bar n \cdot p - \omega \over \bar n \cdot p} \right )   \,.
\end{eqnarray}
To construct the SCET sum rules for the effective form factor $\xi_{\|}(n \cdot p)$,
we write the correlation function as a dispersion integral
\begin{eqnarray}
\Pi_{\nu, \|}(p, q) &=& - {\tilde{f}_B(\mu) \, m_B \over 2}\,
\, \int_0^{+\infty} \, {d \omega^{\prime} \over \omega^{\prime}  - \bar n \cdot p - i0} \,
\left [ \phi_{B, \rm{eff}}^{-}(\omega^{\prime}, \mu)
+ \phi_{B, m}^{+}(\omega^{\prime}, \mu) \right ]  \,  \bar n_{\nu}  \,,
\label{dispersipon form for Pi-L}
\end{eqnarray}
where the effective $B$-meson ``distribution amplitudes" are identical to that used in the $B \to P$ form factors. 
After applying the NLL resummation to the decay constants $\tilde{f}_B$, we derive
\begin{eqnarray}
\xi_{\|}(n \cdot p) &=& 2 \, {U_2(\mu_{h2}, \mu) \, \tilde{f}_B(\mu_{h2})  \over f_{V, \|} } \,
{ m_B \, m_V \over (n \cdot p)^2} \, \nonumber \\
&&  \times \, \int_{0}^{\omega^{V}_s} \, d \omega^{\prime} \,
{\rm exp} \left [ - {n \cdot p \, \omega^{\prime} - m_V^2 \over n \cdot p \, \omega^{V}_M} \right ] \,
\left [ \phi_{B, \rm{eff}}^{-}(\omega^{\prime}, \mu) +  \phi_{B, m}^{+}(\omega^{\prime}, \mu)\right ] \,.
\label{SCET sum rules for xi-L}
\end{eqnarray}
The calculations of the other $\xi_{\perp}$, $\Xi_{\|}$ and $\Xi_{\perp}$ are similar but lengthy, see \cite{Gao:2019lta} for the details.
We will not show the results of these three form factors here.

\section{Power corrections to the form factors}
\label{sec:nlp}

\begin{figure}
\begin{center}
\includegraphics[width=0.35 \columnwidth]{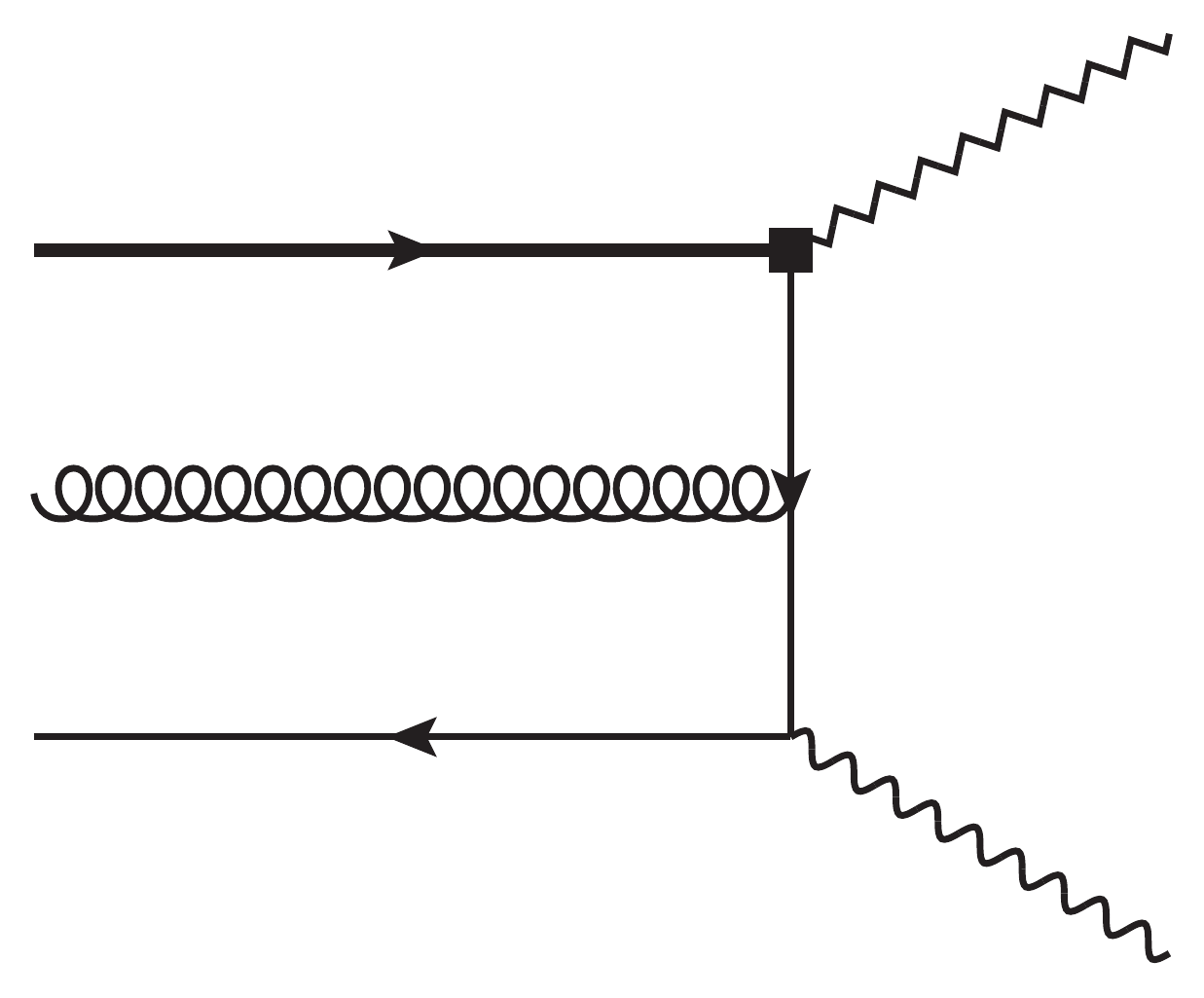}
\vspace*{0.1cm}
\caption{Diagrammatical representation of the three-particle higher-twist corrections
to the vacuum-to-$B$-meson correlation function .
The square box indicates the insertion of the weak vertex $\bar{q} \, \Gamma_\mu  \, b$,
and the waveline represents the interpolating current
$\bar{d} \not \! \! n \, \gamma_5 \, q$ for the light-pseudoscalar meson.  }
\label{fig: correlator for 3P at LO}
\end{center}
\end{figure}

In the previous sections the form factors are calculated at leading power. The power suppressed contributions are is expected to be sizeable due to the finite bottom quark mass. So far, the high twist contribution has been considered in various works. In this section we will take the scalar and vector $B \to P$ form factors as an example, investigate the contribution from
both the two-particle and three-particle $B$-meson LCDAs employing a complete parametrization
of the corresponding three-particle light-cone matrix element and the equation of motion (EOM) constraints of
the higher-twist LCDAs presented in  \cite{Braun:2017liq}.
To obtain the result from three-particle LCDAs, we make use of the light-cone expansion of the quark propagator
in the background gluon field \cite{Balitsky:1987bk}
\begin{eqnarray}
\langle 0 | {\rm T} \, \{\bar q (x), q(0) \} | 0\rangle
 \supset   i \, g_s \, \int_0^{\infty} \,\, {d^4 k \over (2 \pi)^4} \, e^{- i \, k \cdot x} \,
\int_0^1 \, d u \, \left  [ {u \, x_{\mu} \, \gamma_{\nu} \over k^2 - m_q^2}
 - \frac{(\not \! k + m_q) \, \sigma_{\mu \nu}}{2 \, (k^2 - m_q^2)^2}  \right ]
\, G^{\mu \nu}(u \, x) \,, \hspace{0.4 cm}
\end{eqnarray}
where we only keep the one-gluon part without the covariant derivative of the $G_{\mu \nu}$  terms.
The tree-level diagram is  displayed in figure \ref{fig: correlator for 3P at LO},
and  three-particle higher-twist corrections to the vacuum-to-$B$-meson
correlation function can be derived directly

\begin{eqnarray}
\Pi_{\mu}^{(3P)}(n \cdot p,\bar n \cdot p) &=&
-{\tilde{f}_B(\mu) \, m_B \over n \cdot p } \, \int_0^{\infty} \, d \omega_1  \, \int_0^{\infty} \, d \omega_2 \,
\int_0^1 d u \, {1 \over \left [\bar n \cdot p - \omega_1 - u \, \omega_2 \right ]^2} \nonumber \\
&&  \times \, \bigg \{ \bar n_{\mu} \, \left [ \rho_{\bar n, \rm{LP}}^{(3P)}(u, \omega_1, \omega_2, \mu)
+ {m_q \over n \cdot p} \, \rho_{\bar n, \rm{NLP}}^{(3P)}(u, \omega_1, \omega_2, \mu) \right ] \nonumber \\
&& +\,  n_{\mu} \, \left [ \rho_{n, \rm{LP}}^{(3P)}(u, \omega_1, \omega_2, \mu)
+ {m_q \over n \cdot p} \, \rho_{n, \rm{NLP}}^{(3P)}(u, \omega_1, \omega_2, \mu) \right ] \bigg \}   \,,
\end{eqnarray}
where  light-quark mass dependent term of the three-particle corrections have been taken into account, which are suppressed by one power of $\Lambda/m_b$
The explicit expressions of $\rho_{i, \rm{LP}}^{(3P)}$ and $\rho_{i, \rm{NLP}}^{(3P)}$
($i=n\,, \bar n $) can be found in \cite{Lu:2018cfc}.
The definition of three particle LCDAs in \cite{Braun:2017liq} is essential in evaluating these results. Due to nonvanishing quark transverse momentum, the higher-twist two-particle $B$-meson LCDAs are related expressed in terms of the three-particle
configurations with the exact EOM, and they must be taken into account simultaneously
for consistency. The calculation of the higher-twist two-particle $B$-meson LCDAs is quite similar to the leading power calculation. Adding up the two-particle and three-particle higher-twist corrections at tree level together and implementing
the standard strategy to construct the sum rules for heavy-to-light form factors gives rise to  the following
expressions
\begin{eqnarray}
&& {f_{P} \, n \cdot p \over 2} \,\, {\rm exp} \left [- {m_{P}^2 \over n \cdot p \,\, \omega^{P}_M} \right ] \,\,
\left [  f_{B \to P}^{+, \, \rm HT}(q^2)  + \frac{m_B} {n \cdot p} \, f_{B \to P}^{0, \, \rm HT}(q^2)  \right ] \, \nonumber \\
&& = - {\tilde{f}_B(\mu) \, m_B \over n \cdot p}  \,
\bigg \{ e^{-\omega^{P}_s/\omega^{P}_M} \, H_{\bar n, \rm LP}^{\rm 2PHT}(\omega^{P}_s, \mu)
+  \int_0^{\omega^{P}_s} \, d \omega^{\prime}  \, {1 \over \omega^{P}_M} \,
e^{-\omega^{\prime}/\omega_M}  \, H_{\bar n, \rm LP}^{\rm 2PHT}(\omega^{\prime}, \mu) \nonumber \\
&& \hspace{0.4 cm} + \int_0^{\omega^{P}_s} \, d \omega_1 \, \int_{\omega^{P}_s - \omega_1}^{\infty} \, {d \omega_2 \over \omega_2} \,
e^{-\omega^{P}_s/\omega^{P}_M} \,
\bigg [ H_{\bar n, \rm LP}^{\rm 3PHT} \left ({\omega^{P}_s - \omega_1 \over \omega_2}, \omega_1, \omega_2, \mu \right ) \nonumber \\
&& \hspace{0.8 cm} + {m_q \over n \cdot p} \,
H_{\bar n, \rm NLP}^{\rm 3PHT} \left ({\omega^{P}_s - \omega_1 \over \omega_2}, \omega_1, \omega_2, \mu \right ) \bigg ] \nonumber \\
&& \hspace{0.4 cm} + \int_0^{\omega^{P}_s} \, d \omega^{\prime} \, \int_0^{\omega^{\prime}} \, d \omega_1 \,
\int_{\omega^{\prime}  - \omega_1}^{\infty} \, {d \omega_2 \over \omega_2} \, {1 \over \omega^{P}_M} \, e^{-\omega^{\prime}/\omega^{P}_M} \,
\bigg [ H_{\bar n, \rm LP}^{\rm 3PHT} \left ({\omega^{\prime} - \omega_1 \over \omega_2}, \omega_1, \omega_2, \mu \right )
\nonumber \\
&& \hspace{0.8 cm} + {m_q \over n \cdot p} \,H_{\bar n, \rm NLP}^{\rm 3PHT}
\left ({\omega^{\prime} - \omega_1 \over \omega_2}, \omega_1, \omega_2, \mu \right ) \bigg ] \bigg \} \,,
\label{higher twist of fplus}
\\
&& {f_{P} \, n \cdot p \over 2} \,\, {\rm exp} \left [- {m_{P}^2 \over n \cdot p \,\, \omega^{P}_M} \right ] \,
{m_B \over n \cdot p -m_B} \,
\left [  f_{B \to P}^{+, \, \rm HT}(q^2)  - \frac{m_B} {n \cdot p} \, f_{B \to P}^{0, \, \rm HT}(q^2)  \right ] \, \nonumber \\
&& = - {\tilde{f}_B(\mu) \, m_B \over n \cdot p}  \,
\bigg \{  \int_0^{\omega^{P}_s} \, d \omega_1 \, \int_{\omega^{P}_s - \omega_1}^{\infty} \, {d \omega_2 \over \omega_2} \,
e^{-\omega^{P}_s/\omega^{P}_M} \,
\bigg [ H_{n, \rm LP}^{\rm 3PHT} \left ({\omega^{P}_s - \omega_1 \over \omega_2}, \omega_1, \omega_2, \mu \right ) \nonumber \\
&& \hspace{0.8 cm} + {m_q \over n \cdot p} \,
H_{n, \rm NLP}^{\rm 3PHT} \left ({\omega^{P}_s - \omega_1 \over \omega_2}, \omega_1, \omega_2, \mu \right ) \bigg ] \nonumber \\
&& \hspace{0.4 cm} + \int_0^{\omega^{P}_s} \, d \omega^{\prime} \, \int_0^{\omega^{\prime}} \, d \omega_1 \,
\int_{\omega^{\prime}  - \omega_1}^{\infty} \, {d \omega_2 \over \omega_2} \, {1 \over \omega_M} \, e^{-\omega^{\prime}/\omega^{P}_M} \,
\bigg [ H_{n, \rm LP}^{\rm 3PHT} \left ({\omega^{\prime} - \omega_1 \over \omega_2}, \omega_1, \omega_2, \mu \right )
\nonumber \\
&& \hspace{0.8 cm} + {m_q \over n \cdot p} \,H_{n, \rm NLP}^{\rm 3PHT}
\left ({\omega^{\prime} - \omega_1 \over \omega_2}, \omega_1, \omega_2, \mu \right ) \bigg ]  \bigg \} \,,
\label{higher twist of fzero}
\end{eqnarray}
where the nonvanishing spectral functions  $H_{i, \rm LP}^{\rm 2PHT}$ and  $H_{i, \rm (N)LP}^{\rm 3PHT}$
($i = n, \, \bar n$) are given by
\begin{eqnarray}
H_{\bar n, \rm LP}^{\rm 2PHT}(\omega, \mu) &=&
4 \, \hat{g}_B^{-}(\omega, \mu) \,, \nonumber \\
 H_{n, \rm LP}^{\rm 3PHT} (u, \omega_1, \omega_2, \mu)&=& 2 \, (u-1) \, \phi_4(\omega_1, \omega_2, \mu)  \,, \nonumber \\
H_{n, \rm NLP}^{\rm 3PHT} (u, \omega_1, \omega_2, \mu)&=& \tilde{\psi}_5(\omega_1, \omega_2, \mu)
- \psi_5(\omega_1, \omega_2, \mu)  \,, \nonumber \\
H_{\bar n, \rm LP}^{\rm 3PHT} (u, \omega_1, \omega_2, \mu)&=& \tilde{\psi}_5(\omega_1, \omega_2, \mu)
- \psi_5(\omega_1, \omega_2, \mu)  \,, \nonumber \\
H_{\bar n, \rm NLP}^{\rm 3PHT} (u, \omega_1, \omega_2, \mu)&=&  2 \,\phi_6(\omega_1, \omega_2, \mu)  \,.
\end{eqnarray}
It is evident that the two-particle higher-twist corrections preserve the large-recoil symmetry relations
of the $B \to P$ form factors and the three-particle higher-twist contributions violate such relations
already at tree level .
Employing the power counting scheme for the Borel mass $\omega^{P}_M$ and the threshold parameter $\omega^{P}_s$ \cite{Wang:2015vgv}
\begin{eqnarray}
\omega^{P}_s \sim \omega^{P}_M  \sim {\cal O}(\Lambda^2/m_b)\,,
\end{eqnarray}
we can identify the scaling behaviours of the higher-twist corrections to $B \to P$ form factors
\begin{eqnarray}
f_{B \to P}^{+, \, \rm HT}(q^2) \sim f_{B\to  P}^{0, \, \rm HT}(q^2) \sim
f_{B \to P}^{T, \, \rm HT}(q^2) \sim {\cal O}  \left ( {\Lambda \over m_b} \right )^{5/2}  \,
\end{eqnarray}
in the heavy quark limit, which is suppressed by one power of $\Lambda/m_b$ compared with the leading-twist contribution. Collecting different pieces together, the final expressions for the LCSR of $B \to P$ form factors at large hadronic recoil can be written as
\begin{eqnarray}
f_{B \to P}^{+}(q^2) &=&  f_{B \to P}^{+, \rm 2PNLL}(q^2) + f_{B \to P}^{+, \, \rm 2PHT}(q^2)
+ f_{B \to P}^{+, \, \rm 3PHT}(q^2)  \,, \nonumber \\
f_{B \to P}^{0}(q^2) &=&  f_{B \to P}^{0, \rm 2PNLL}(q^2) + f_{B \to P}^{0, \, \rm 2PHT}(q^2)
+ f_{B \to P}^{0, \, \rm 3PHT}(q^2)  \,,
\label{final sum rules}
\end{eqnarray}
where the manifest expressions of $f_{B \to P}^{i, \rm 2PNLL}(q^2)$ ($i=+, \, 0$) includes the light-quark mass effect, and the higher-twist corrections  $f_{B \to P}^{i, \, \rm 2PHT}(q^2)$ and $f_{B \to P}^{i, \, \rm 3PHT}(q^2)$ can be extracted from
(\ref{higher twist of fplus}), (\ref{higher twist of fzero}).

The power suppressed contribution is from much more sources than the high twist contribution considered here. Various kinds of power suppressed contribution  to the $ B \to \gamma \nu \ell$ \cite{Wang:2016qii, Wang:2018wfj, Beneke:2018wjp} and $B_s \to \gamma\gamma$ \cite{Shen:2020hfq} has been comprehensively studied, among which some kinds of sources can also contribute to the heavy-to-light form factor. One example is from heavy quark expansion, i.e. 
\begin{equation}
\begin{aligned}		
\bar q \, \Gamma_\mu \, b  = e^{-im_b v\cdot x} \, \bar q \, \Gamma_\mu \left [ 1 + {i \vec{\slashed D} \over 2m_b} \right ] \, h_v+ \cdots ,
\end{aligned} \label{b-quark-expansion}  \end{equation}\\
where $\vec{D}^\mu=D^\mu-(v\cdot D) \, v^\mu $, and the $b$-quark mass $m_b$ is defined in the pole-mass scheme in HQET. It is clear replacing the first term in the square bracket with the secon term yields  power suppressed contributions.  In addition, the power suppressed local and nonlocal  contribution in hard-collinear propagators might also play an important role. A specific calculation of these contributions will be published elsewhere.

\section{Numerical results}
\label{sec:results}

Have the sum rules for heavy-to-light form factors at hand, we will proceed to perform the numerical analysis.
First we will show the predictions of the form factors with LCSR at large recoil.
Then we will explore some phenomenological implications of the form factors by extrapolating the form factors to the entire physical region with the $z$-series parametrization.

\subsection{LCSR predictions of the form factors}

\begin{table}[!t]
\centering
\renewcommand{\arraystretch}{1.6}
\resizebox{\columnwidth}{!}{
\begin{tabular}{|l|ll||l|ll|}
\hline
  Parameter
& Value
& Ref.
&  Parameter
& Value
& Ref.
\\
\hline
$m_{s}(2\, {\rm GeV})$    & $93.8 \pm 1.5 \pm 1.9 \,\, {\rm MeV} $
& \cite{Tanabashi:2018oca}
& $m_{b}(m_{b})$    & $4.193^{+0.022}_{-0.035} \,\, {\rm GeV} $
& \cite{Beneke:2014pta,Dehnadi:2015fra} 
\\
$\mu$ & $1.5\pm 0.5~ {\rm GeV}$ & ~
& $\mu_{h1},\mu_{h2},\nu_{h}$ &
${m_{b}}^{+m_{b}}_{-m_{b}/2}$ & ~
\\
\hline
$f_{B}$ & $192.0 \pm 4.3 \,\, {\rm MeV} $ & \cite{Aoki:2016frl}
&  & &
\\
$f_{\pi}$ & $130.2 \pm 1.7 \,\, {\rm MeV} $ & \cite{Tanabashi:2018oca}
& $f_{K}$ & $155.6 \pm 0.4 \,\, {\rm MeV} $ & \cite{Tanabashi:2018oca}
\\
$f_{\rho,\|}$ & $213 \pm 5 \,\, {\rm MeV} $ & \cite{Straub:2015ica}
& $f_{\rho,\perp}(1 \,{\rm GeV})$ & $160 \pm 7 \,\, {\rm MeV} $ & \cite{Straub:2015ica}
\\
$f_{\omega,\|}$ & $197 \pm 8 \,\, {\rm MeV} $ & \cite{Straub:2015ica}
& $f_{\omega,\perp}(1 \,{\rm GeV})$ & $148 \pm 13 \,\, {\rm MeV} $ & \cite{Straub:2015ica}
\\
$f_{K^{\ast},\|}$ & $204 \pm 7 \,\, {\rm MeV} $ & \cite{Straub:2015ica}
& $f_{K^{\ast},\perp}(1 \,{\rm GeV})$ & $159 \pm 6 \,\, {\rm MeV} $ & \cite{Straub:2015ica}
\\
\hline
$s^{\pi}_{0}$ & $0.70\pm0.05~ {\rm GeV}^{2}$ &
\cite{Wang:2015vgv,Khodjamirian:2003xk}
& $s^{K}_{0}$ & $1.05\pm0.05~ {\rm GeV}^{2}$ &
\cite{Wang:2015vgv,Khodjamirian:2003xk}
\\
$s^{\rho,\|}_{0}$ & $1.5 \pm 0.1~ {\rm GeV}^{2}$ &
\cite{Ball:1998sk,Khodjamirian:2006st}
& $s^{\rho,\perp}_{0}$ & $1.2\pm0.1~ {\rm GeV}^{2}$ &
\cite{Ball:1996tb}
\\
$s^{\omega,\|}_{0}$ & $s^{\rho,\|}_{0}+ m^2_{\omega} - m^2_{\rho}$ &
\cite{Ball:1998sk,Khodjamirian:2006st}
& $s^{\omega,\perp}_{0}$ & $s^{\rho,\perp}_{0}+ m^2_{\omega} - m^2_{\rho}$ &
\cite{Ball:1996tb}
\\
$s^{K^{\ast},\|}_{0}$ & $s^{\rho,\|}_{0}+ m^2_{K^{\ast}} - m^2_{\rho}$ &
\cite{Ball:1998sk,Khodjamirian:2006st}
& $s^{K^{\ast},\perp}_{0}$ & $s^{\rho,\perp}_{0}+ m^2_{K^{\ast}} - m^2_{\rho}$ &
\cite{Ball:1996tb}
\\
$M^{2}_{\pi,K}$ & $1.25\pm0.25~ {\rm GeV}^{2}$ & \cite{Wang:2015vgv,Khodjamirian:2003xk}
& $M^{2}_{\rho}$ & $1.5\pm0.5~ {\rm GeV}^{2}$ &
\cite{Ball:1998sk}
\\
$M^{2}_{\omega}$ & $M^2_{\rho} + m^2_{\omega} - m^2_{\rho}$ &
\cite{Ball:1998sk}
& $M^{2}_{K^{\ast}}$ & $M^2_{\rho} + m^2_{K^{\ast}} - m^2_{\rho}$ &
\cite{Ball:1998sk}
\\
\hline
\end{tabular}
}
\renewcommand{\arraystretch}{1.0}
\caption{The numerical values of  the various input  parameters employed
in the  theory predictions of the $B\to M$ form factors with $B$-meson LCSR. }
\label{tab:inputs}
\end{table}

Prior to presenting the predictions of
the semileptonic $B \to M$ decay form factors, we need to determine the input parameter used in the numerical analysis.
For the later convenience, we collect the important inputs in table \ref{tab:inputs}.
Except for these parameters, we still need to determine the inverse moment of the $B$ meson $\lambda_B(\mu_0)$.
The parameter $\lambda_B(\mu_0)$ could be calculated with nonperturbative
HQET sum rules \cite{Braun:2003wx} and could also be extracted indirectly from $B \to \gamma \ell \nu$ process
\cite{Ball:2003fq,Wang:2018wfj,Braun:2012kp,Wang:2016qii,Beneke:2018wjp}.
While at present, we are still lack of a satisfied constraints of $\lambda_B(\mu_0)$.
Following the strategy displayed in \cite{Wang:2015vgv}, we will employ a fit approach to determine $\lambda_B(\mu_0)$.

For the $B\to P$ form factors, matching the LCSR calculation for the
vector $B \to \pi$ form factor at $q^2=0$
with the predictions from LCSR with pion LCDAs $f_{B \to \pi}^{+}(q^2=0)=0.28 \pm 0.03$ \cite{Khodjamirian:2011ub,Imsong:2014oqa,Khodjamirian:2017fxg} will provides one with a suitable value of $\lambda_B(\mu_0)$.
While for the $B\to V$ form factors, $\lambda_B(\mu_0)$ could be determined by matching the LCSR prediction of the form factor $V_{B \to \rho}(q^2=0)$ to that of the improved NLO LCSR with the $\rho$-meson LCDAs \cite{Straub:2015ica}.
Performing such matching procedure we obtain
\begin{eqnarray}
\lambda_B(\mu_0) = \left\{
\begin{array}{l}
285^{+27}_{-23} \,\, {\rm MeV},   \qquad  \hspace{1.5 cm}
(\rm B\to P)  \vspace{.3 cm}\\
343^{+22}_{-20} \,\, {\rm MeV},
 \qquad  \hspace{1.5 cm}
(\rm B\to V)
\end{array}
 \hspace{0.5 cm} \right.
\label{values of lambdaB}
\end{eqnarray}
It is interesting to notice that the extracted values of $\lambda_B(\mu_0)$ for the $B\to V$ form factors are in nice agreement with that of the $B\to P$ form factors and are also consistent with the implications of experimental data
for the two-body charmless hadronic $B$-meson decays from the QCD factorization approach \cite{Beneke:2003zv}.

\begin{figure}[t]
\begin{center}
\includegraphics[width=0.55 \columnwidth]{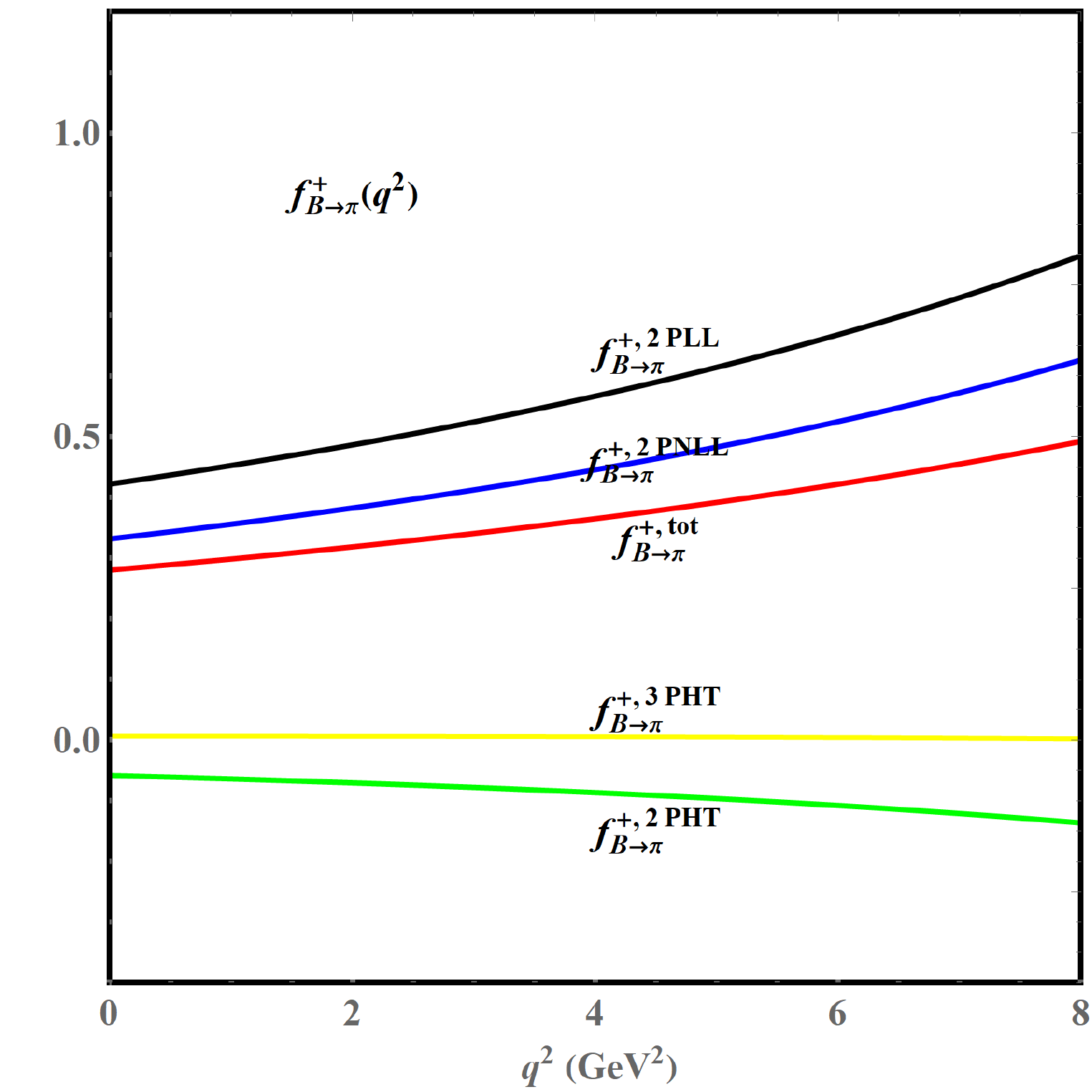}
\vspace*{0.1cm}
\caption{The $q^{2}$ dependence of the form factor $f^{+}_{B\to\pi}(q^{2})$.
The red line represent the total contribution.
The black, blue, green and yellow correspond to the LP contribution at LL, the LP contribution at NLL, the two-particle higher-twist correction and the three-particle higher-twist correction, respectively.}
\label{fig: Breakdown of the vector B to pi form factor}
\end{center}
\end{figure}

Employing above-determined parameters, one can explore the LCSR predictions of the $B\to M$ form factors in large-recoil region.
As an illustration, we show the breakdown of different contributions to the form factor $f^{+}_{B\to\pi}(q^{2})$
in figure \ref{fig: Breakdown of the vector B to pi form factor}.
One can see from the figure that higher-twist corrections to the form factor
$f_{B \to \pi}^{+}(q^2)$ are dominated by the two-particle twist-five $B$-meson LCDA $\hat{g}_B^{-}(\omega, \mu)$,
which can shift the leading-power prediction by an amount of approximately $(20 \sim 30) \%$.
Compared the the two-particle higher-twist contribution, the three-particle higher-twist contribution only generates a tiny contribution on the form factor.
We further find that the radiative correction can introduce about a ${\cal O}(20 \%)$ reduction of the corresponding LL prediction.
We have also verified that such observations also hold true for the other $B \to M$ form factors at large hadronic recoil.

\subsection{Phenomenological applications}

It is well-known that the light-cone operator product expansion of the vacuum-to-$B$-meson correlation function is only valid in the large-recoil region \cite{Khodjamirian:2006st,Wang:2015vgv}.
One then need to extrapolate the LCSR predictions of $B \to M$ form factors at $q^2 \leq 8 \, {\rm GeV}^2$
to the full kinematic region with the $z$-series expansion.
The parameter $z$ corresponds a map of the entire cut $q^2$-plane onto a unit disk $|z(q^2, \, t_0)|\leq 1$
\begin{eqnarray}
z(q^2, t_0) = \frac{\sqrt{t_{+}-q^2}-\sqrt{t_{+}-t_0}}
{\sqrt{t_{+}-q^2}+\sqrt{t_{+}-t_0}}  \,.
\end{eqnarray}
Here, $t_{+} = (m_B + m_{M})^2$ is determined by the threshold of the lowest continuum state
which can be generated by the weak transition currents in QCD.
The auxiliary parameter $t_0$, which determines the $q^2$ point to be mapped onto the origin
of the complex $z$ plane, will be further choosen as \cite{Khodjamirian:2017fxg,Lu:2018cfc}
\begin{align}
& t_0 = (m_B + m_P) \, (\sqrt{m_B} + \sqrt{m_P})^2 \,,
&& B\to P
\nonumber \\
&t_0=(m_B - m_V)^2\,.
&& B\to V
\end{align}
We will use the simplified Bourrely-Caprini-Lellouch (BCL) series expansion for the $B \to M$ form factors
\cite{Bourrely:2008za} (see \cite{Boyd:1994tt,Boyd:1995cf} for an alternative parametrization and \cite{Khodjamirian:2011ub} for more discussions
for the $B \to \pi$ form factors)
\begin{eqnarray}
f_{B \to P}^{+, T}(q^2) &=&  {f_{B \to P}^{+, T}(0) \over 1 - q^2/m_{B_{(s)}^{\ast}}^2} \,
\bigg \{ 1 + \, \sum_{k=1}^{N-1}   \, b_{k, P}^{+, T}  \,
\bigg  ( z(q^2, \, t_0)^k -  z(0, \, t_0)^k  \nonumber \\
&& - \, (-1)^{N-k} \, {k \over N} \,
\left [  z(q^2, \, t_0)^N -  z(0, \, t_0)^N \right ]  \bigg  ) \bigg \}   \,,
\nonumber \\
f_{B \to P}^{0}(q^2) &=&  f_{B \to P}^{0}(0) \,
\left \{ 1 +  \, \sum_{k=1}^{N}   \, b_{k, P}^{0}  \,
\left (  z(q^2, \, t_0)^k -  z(0, \, t_0)^k  \right )   \right \}   \,,
\nonumber \\
f_{B \to V}^{i}(q^2) &=& {f_{B \to V}^{i}(0) \over 1-q^2/m_{i, \, \rm pole}^2} \,
\left \{1 + \sum_{k=1}^N \, b_{k,V}^i \, \left [z(q^2, t_0)^k - z(0, t_0)^k \right ]  \right \}  \,.
\label{z expansion of B to M FFs}
\end{eqnarray}
The expansion parameter is small in the entire region $|z(q^2, t_0)|^2 \leq 0.04$, the expansion could be truncated at certain $N$.
We will truncate the $z$-series at $N=2$  for the form factors $f_{B \to P}^{+, T}$
and at $N=1$ for the all the other form factors
(see \cite{Bharucha:2010im} for further  discussions on the systematic truncation uncertainties).
The adopted values of the various resonance masses from the Particle Data Group (PDG) \cite{Tanabashi:2018oca}
and from the heavy-hadron chiral perturbation theory \cite{Bardeen:2003kt} are summarized
in table \ref{table: inputs of resonance masses}.

\begin{table}[t!bph]
\renewcommand{\arraystretch}{2.}
\begin{center}
\begin{tabular}{|c|c|c|c|}
  \hline
  $f_{B \to V}^{i}(q^2)$ & \,\,\, $J^P$ \,\,\, & $b \to d$ \, (in  ${\rm GeV}$) & $b \to s$ \, (in ${\rm GeV}$) \\
  \hline
 $V(q^2)$, \, $T_1 (q^2)$  & $1^{-}$ & 5.325 & 5.415 \\
 $A_0 (q^2)$ & $0^{-}$ & 5.279 & 5.366 \\
 $A_1 (q^2)$,  \, $A_{12} (q^2)$,  \, $T_2 (q^2)$, \, $T_{23}(q^2)$
 & $1^{+}$ & 5.724 & 5.829 \\
  \hline
\end{tabular}
\end{center}
\caption{The resonance masses with different quantum numbers entering the $z$-series
expansions of the $B \to V$ form factors (\ref{z expansion of B to M FFs}) where $A_{12}={m_B + m_V \over n \cdot p} \, A_1
- {m_B - m_V \over m_B} \, A_2$ and $T_{23}=\frac{m_{B}}{n\cdot p}T_{2}-T_{3}$.}
\label{table: inputs of resonance masses}
\end{table}

Having at our disposal the theory predictions for $B \to \pi$ form factors,
we proceed to explore  phenomenological aspects of the semileptonic $B \to \pi \ell \nu$  decays,
which serves as the golden channel for the determination of CKM matrix element $|V_{ub}|$ exclusively
(see \cite{Kou:2018nap} for the future advances of precision measurements of Belle II).
It is straightforward to write down the differential decay rate for $B \to \pi \ell \bar \nu_{\ell}$
\begin{eqnarray}
{d \,  \Gamma (B \to \pi \ell \bar \nu_{\ell})\over d q^2}
&=& {G_F^2 \, |V_{ub}|^2 \over 192 \, \pi^3 \, m_B^3} \, \lambda^{3/2}(m_B^2, m_{\pi}^2, q^2) \,
\left (  1 -{m_l^2 \over  q^2} \right )^2  \,  \left (  1 + {m_l^2 \over  2 \, q^2} \right ) \,
\bigg [ |f_{B \to \pi}^{+} (q^2)|^2  \nonumber \\
&& +  \, {3 \, m_l^2 \, (m_B^2 - m_{\pi}^2)^2 \over \lambda(m_B^2, m_{\pi}^2, q^2) \, (m_l^2 + 2\, q^2) } \,
|f_{B \to \pi}^{0} (q^2)|^2   \bigg ]  \,,
\nonumber \\
 \frac{d \Gamma(B \to V \, \ell \, \bar \nu_{\ell})}{d q^2 } &=&
{G_F^2 \,|V_{ub}|^2 \over 192 \, \pi^3 \, m_B^3} \, {q^2 \over c_V^2} \, \lambda^{1/2}(m_B^2, m_V^2, q^2) \,
\bigg \{|H_0(q^2)|^2
+ |H_{+}(q^2)|^2 +|H_{-}(q^2)|^2  \bigg \}  \,,
\end{eqnarray}
where $\lambda(a, b, c) = a^2 + b^2 + c^2- 2 \, ab -2 \, ac - 2 \, bc$ and  the three  helicity amplitudes $H_{i}(q^2)$ ($i=\pm, \, 0$) can be expressed in terms of the semileptonic
$B \to V$ form factors
\begin{eqnarray}
H_{\pm}(q^2) &=& (m_B +m_V) \, \left [ A_1(q^2) \mp { 2\, m_B \, |\vec{p}_{V}| \over (m_B+m_V)^2} \, V(q^2)\right ] \,,
\nonumber  \\
H_{0}(q^2) &=&  {m_B +m_V \over 2 \, m_V \, \sqrt{q^2}} \,
\bigg [ (m_B^2 - m_V^2 -q^2) \, A_1(q^2)
- {4 \, m_B^2 \, |\vec{p}_{V}|^2 \over (m_B +m_V)^2} \, A_2(q^2) \bigg ]   \,,
\end{eqnarray}
with the momentum $|\vec{p}_{V}|$ of the light-vector meson in the $B$-meson rest frame given by
$
|\vec{p}_{V}| =  {1 \over 2 \, m_B} \, \lambda^{1/2}(m_B^2, m_V^2, q^2)$.

Employing the experimental measurements
of $B \to \pi \ell \bar\nu_{\ell}$ \cite{Lees:2012vv,Sibidanov:2013rkk},
and taking advantage of the measurements of the partial branching fractions for $B \to \rho \, \ell \, \bar \nu_{\ell}$
\cite{delAmoSanchez:2010af,Sibidanov:2013rkk} and $B \to \omega \, \ell \, \bar \nu_{\ell}$ \cite{Lees:2012vv,Sibidanov:2013rkk}
we can derive the following intervals for exclusive $|V_{ub}|$
\begin{eqnarray}
|V_{ub}|_{\rm exc.} = \bigg ( 3.23\,{}^{+0.66}_{-0.48} \big |_{\rm th.}\,{}^{+0.11}_{-0.11} \big |_{\rm exp.} \bigg )
\times 10^{-3} \,,
\qquad  [{\rm from} \,\, B \to \pi \ell \nu_{\ell} ]
\nonumber \\
|V_{ub}|_{\rm exc.} = \bigg ( 3.05\,{}^{+0.67}_{-0.52} \big |_{\rm th.}\,{}^{+0.19}_{-0.20} \big |_{\rm exp.} \bigg )
\times 10^{-3} \,,  \qquad  [{\rm from} \,\, B \to \rho \ell \nu_{\ell} ] \nonumber \\
|V_{ub}|_{\rm exc.} = \bigg ( 2.54 \,{}^{+0.56}_{-0.40} \big |_{\rm th.}\,{}^{+0.18}_{-0.19} \big |_{\rm exp.} \bigg )
\times 10^{-3} \,.  \qquad  [{\rm from} \,\, B \to \omega \ell \nu_{\ell} ]
\end{eqnarray}
Apparently,  the extracted values of $|V_{ub}|$ from
$B \to V \, \ell \, \bar \nu_{\ell}$ decay are lower than that from the $B \to \pi \, \ell \, \bar \nu_{\ell}$ channel.
The values of $|V_{ub}|$ from $B \to \pi \, \ell \, \bar \nu_{\ell}$ decay are in agreement with the averaged exclusive determinations presented in PDG \cite{Tanabashi:2018oca},
while the central values of both determinations of $|V_{ub}|$ from $B \to V \, \ell \, \bar \nu_{\ell}$
are somewhat smaller than the corresponding result in PDG \cite{Tanabashi:2018oca}.
We also observe that the obtained $|V_{ub}|$ from
the $B \to \omega \, \ell \, \bar \nu_{\ell}$ process
are significantly smaller than that from the exclusive channel  $B \to \rho \, \ell \, \bar \nu_{\ell}$ as already observed in
\cite{Sibidanov:2013rkk}.
All the three exclusively extracted values of $|V_{ub}|$ from
the $B \to M \, \ell \, \bar \nu_{\ell}$ decays are significantly smaller than the
averaged inclusive determinations reported in \cite{Tanabashi:2018oca}
\begin{eqnarray}
|V_{ub}|_{\rm inc.} = \bigg ( 4.49 \pm 0.15 \,{}^{+0.16}_{-0.17}  \pm 0.17 \bigg )
\times 10^{-3} \,.
\end{eqnarray}

\section{Summary}

In this review we have discussed the LCSR with $B$-meson LCDAs and its application to calculating the $B\to P$ and $B \to V$ form factors.  The fundamental nonperturbative inputs in this approach are the LCDAs of the $B$ meson, which are defined in terms of nonlocal operators including the effective $b$-quark field and the soft light fields sandwiched between the vacuum and the $B$-meson state. At the one-loop level, the leading-twist $B$-meson LCDA satisfies the Lange-Neubert equation, which can be simplified by some kinds of integral transforms. The two-loop level evolution equation of the leading-twist $B$-meson LCDA has also been obtained. The evolution function is useful in the determination of the models of the LCDAs, especially the behavior at the endpoint. The higher-twist $B$-meson LCDAs can be introduced by adjusting the spinor structure, or adding an additional gluon field, or taking the higher power contribution from the light-cone expansion. They can be studied using a similar method as the leading-twist one.

We introduced the investigation of the $B\to M$ form factors with the $B$-meson LCSR in detail. To obtain the sum rules, one must start from the correlation function which is defined as the vacuum-to-$B$-meson matrix element of the time-ordered product of the weak current and the interpolation current of the light meson. At the partonic level, the correlation function can be factorized into the convolution of the hard function, the jet function, and the LCDAs of the $B$ meson. The short distance hard function and jet function can be calculated by choosing free partonic external states. To evaluate the NLO corrections to the hard function and the jet function, one might employ the method of regions, i.e., the loop momentum was assigned to be hard, collinear, and soft,  respectively. 
The loop integral was calculated in different regions, and the hard function and jet function can be extracted directly. An alternative approach to obtain the NLO hard function and jet function is to perform the match from QCD first to SCET$_{\rm I}$ and then to SCET$_{\rm II}$ respectively. The most subtle place in the loop calculation is the infrared subtraction, especially under the condition that the evanescent operators took the place. The convolution of the hard function, the jet function, and the LCDAs is independent of the factorization scale. However, the large logarithmic terms need to be resummed with the RG equation approach. Having the hard function and the jet function at hand, the correlation function can be expressed in terms of the dispersion integral.
After this, we employed the quark-hadron duality assumption and the Borel transformation, and then obtained the sum rules for the form factors.

We have discussed the power suppressed contributions to the form factors.
The higher-twist corrections to the form factors from both the two-particle and three-particle $B$-meson LCDAs were calculated at tree level. For the contribution from three-particle $B$-meson LCDAs, we employed the quark propagator in the background field. The quark-mass dependent term in the sum rules with three-particle $B$-meson LCDAs is power suppressed, which is different from the leading-twist contribution. Except for the high-twist contribution, there exist more sources of power suppressed contributions,  such as the power suppressed contribution from heavy quark expansion, the power suppressed contribution from hard-collinear propagators, the local terms, etc. The investigation of power suppressed contributions is one of the main goals of future studies.

After employing appropriate input parameters, we evaluated the $B \to P$ and $B \to V$ form factors numerically within the momentum region $0<q^2<8{\rm GeV}^2$ where the LCSR with $B$-meson LCSR is applicable. Then we extrapolated the result to the whole physical region by the $z$-series expansion.  Applying the experimental measurements of the $B \to \pi \ell \bar\nu_{\ell}$ process 
and taking advantage of the measurements of the partial branching fractions for $B \to \rho \, \ell \, \bar \nu_{\ell}$  and $B \to \omega \, \ell \, \bar \nu_{\ell}$ decays, we obtained  intervals for exclusive $|V_{ub}|$. The extracted values of $|V_{ub}|$ from $B \to V \, \ell \, \bar \nu_{\ell}$ decay are lower than that from the $B \to \pi \, \ell \, \bar \nu_{\ell}$ channel. The values of $|V_{ub}|$ from $B \to \pi \, \ell \, \bar \nu_{\ell}$ decay are in agreement with the averaged exclusive determinations presented in PDG. All of the three exclusively extracted values of $|V_{ub}|$ from the $B \to M \, \ell \, \bar \nu_{\ell}$ decays are significantly smaller than the
averaged inclusive determinations.

\subsection*{Acknowledgements}
We thank Y. M. Wang for a resultful cooperation and very valuable discussions. Y.L.S acknowledges the Natural Science Foundation of Shandong province with Grant NO. ZR2020MA093 and the National Natural Science Foundation of China with Grant No. 12175218.
Y.B.W is supported in part by the Alexander-von-Humboldt Stiftung.

\bibliography{ref}
\bibliographystyle{h-physrev}
\end{document}